\documentclass[12pt]{article}
\usepackage[margin=1.0in]{geometry}
\usepackage[utf8]{inputenc}
\usepackage{siunitx}
\usepackage{tabularx}

\usepackage{ltablex}
\usepackage{lscape}
\usepackage{booktabs}
\usepackage{array}
\newcolumntype{x}[1]{>{\centering\let\newline\\\arraybackslash\hspace{0pt}}p{#1}}
\usepackage{amsmath}
\usepackage{amsfonts}
\usepackage{xcolor}
\usepackage{chemformula}
\usepackage{soul}
\usepackage{authblk}
\usepackage{parskip}
\newcommand*{\plimsoll}{{\ensuremath{-\kern-5pt{\ominus}\kern-5pt-}}}
\usepackage[normalem]{ulem}
\usepackage{svg}
\usepackage{bm}
\usepackage{makecell}
\usepackage{placeins}
\usepackage{subcaption}
\usepackage{setspace}

\usepackage[style=numeric,sorting=none,bibstyle=ama,style=numeric-comp,maxbibnames=99]{biblatex} 
\DeclareUnicodeCharacter{2212}{-}
\DeclareUnicodeCharacter{0300}{`}

\newcommand{\beginsupplement}{%
        \setcounter{table}{0}
        \renewcommand{\thetable}{S\arabic{table}}%
        \setcounter{figure}{0}
        \renewcommand{\thefigure}{S\arabic{figure}}
        \setcounter{equation}{0}
        \renewcommand{\theequation}{S\arabic{equation}}%
     }

\title{Dynamic Optimization of Proton Exchange Membrane Water Electrolzyers Considering Usage-Based Degradation}
\author[1]{Landon Schofield}
\author[2]{Benjamin Paren}
\author[3]{Ruaridh Macdonald}
\author[4]{Yang Shao-Horn}
\author[5]{Dharik Mallapragada}
\date{}
\affil[1]{Department of Chemical Engineering, Massachusetts Institute of Technology}
\affil[2]{Department of Chemical Engineering and Materials Science, Stevens Institute of Technology}
\affil[3]{MIT Energy Initiative}
\affil[4]{Department of Mechanical Engineering, Massachusetts Institute of Technology}
\affil[5]{Department of Chemical and Biomolecular Engineering, New York University}

\begin{document}
\begin{refsection}
\maketitle
\begin{abstract}
We present a techno-economic optimization model for evaluating the design and operation of proton exchange membrane (PEM) electrolyzers, crucial for hydrogen production powered by variable renewable electricity. This model integrates a 0-D physics representation of the electrolyzer stack, complete mass and energy balances, operational constraints, and empirical data on use-dependent degradation. Utilizing a decomposition approach, the model predicts optimal electrolyzer size, operation, and necessary hydrogen storage to satisfy baseload demands across various technology and electricity price scenarios. Analysis for 2022 shows that including degradation effects raises the levelized cost of hydrogen from \$4.56/kg to \$6.60/kg and decreases stack life to two years. However, projections for 2030 anticipate a significant reduction in costs to approximately \$2.50/kg due to lower capital expenses, leading to larger stacks, extended lifetimes, and less hydrogen storage. This approach is adaptable to other electrochemical systems relevant for decarbonization.
\end{abstract}

\doublespacing
\section{Introduction}
Policies emphasizing economy-wide decarbonization are increasingly focused on promoting production of low-carbon hydrogen (H$_2$) to address emissions from difficult-to-electrify end-uses. For example, the emissions-indexed production tax credit (PTC) for \ch{H2} as part of the Inflation Reduction Act in the U.S. provides up to \$3/kg \ch{H2}, which has led to several project announcements for H$_2$ production using electrolyzers powered by low-carbon electricity \cite{Yarmuth2021-wu,UnknownUnknown-ci,Arjona2023-bc}.  Proposed electrolyzer projects include both “islanded” configurations, where renewable electricity is co-located with the electrolyzer and is the sole source of electricity input, as well as “grid-connected” systems involving using grid electricity (typically from the wholesale market) as well as on-site or contracted (via power purchase agreements) renewable electricity supply. In both cases, cost-effective operation of the electrolyzer will likely involve operating at much less than 100\% capacity utilization to manage fluctuations in variable renewable energy (VRE) availability as well as grid electricity prices (that are also increasingly influenced by VRE availability) \cite{Stansberry2017-eo,Laoun2016-mb,Sarrias-Mena2015-vy,Saadi2016-xg,Zhao2015-bk}. This part load operation not only has implications for capital utilization, which has been extensively studied, but also stack lifetime through accelerated degradation which is less well studied and could impact the overall economics of H$_2$ production \cite{Kalinci2015-re,Shibata2015-qt,Hurtubia2021-jo}. Here, we focus on the answering this question in the context of grid-connected proton exchange membrane (PEM) electrolyzers.

As of 2022, global installations of electrolyzers totalled 700 MWe \footnote {Unlike many technologies, electrolyzer capacity is typically reported on the basis of nameplate electricity input rather than \ch{H2} output}, mostly composed of alkaline electrolyzers \cite{International_Energy_Agency2023-wq}. While the share of PEM electrolyzers is relatively small at 30\% of total installed capacity, several appealing attributes compared to alkaline systems are expected to drive their deployment in the future \cite{International_Energy_Agency2023-wq}. In particular: a) their ability to operate at higher current densities ($\geq$ 1 \si{\ampere\per\square \centi\meter}), which allows for smaller stack areas (and possibly lower areal footprint) to produce the same amount of H$_2$, and b) ability to operate with a differential pressure between the anode and cathode, which enables production of high pressure H$_2$ product \cite{Carmo2013-bq, International_Energy_Agency2023-wq}. The greater range of current density operation also creates the potential for increased operational flexibility which is an important criteria for cost-effective electrolyzer projects to manage fluctuations in temporal attributes, such as emissions or cost, of electricity supply \cite{Guerra2019-ua,Corengia2022-ha,Chung2024-vr,Tsay2023-se}. The importance of such operational flexibility for power systems balancing is likely to grow with an increasing share of VRE supply in future low-carbon grids, owing increasing instances of near-zero prices and very high prices compared to today's fossil-fuel dominant systems \cite{Mallapragada_undated-iq}.

Several techno-economic optimization studies have highlighted the importance of dynamic operation in minimizing the operating cost of other electrochemical processes when utilizing electricity sourced from VRE that is co-located or supplied via connection to the electric grid \cite{Otashu2019-ge,Hofmann2022-ti,Weigert2021-ik}. Studies on PEM systems reveal that cost-optimal dynamic operation involves operating at high current densities (say $\gg$ 2 \si{\ampere\per\square \centi\meter}) during times of abundant VRE supply or low electricity prices, as well as periods of idling or near-zero current densities during high electricity prices or low VRE supply periods, that collectively results in smaller stack areas when compared to an equivalent steady state electrolyzer operating at current densities near  1-2 \si{\ampere\per\square \centi\meter} \cite{Espinosa-Lopez2018-hb,Stansberry2020-zm,Chung2024-vr,Corengia2020-on,Corengia2022-ha}.

Dynamic operation of PEM electrolyzers via modulation of the operating current density brings with it several challenges -- namely, heat management, gas crossover and associated safety concerns, and stack degradation. Heat management in PEM systems is typically undertaken via excess feed water on the anode side \cite{Stansberry2020-zm,Holst2021-nd}. In an effort to model a 46 kWe electrolyzer connected to renewable energy, Espinosa et. al developed a model of the temperature response of PEM systems at varying current density with this excess feed water strategy \cite{Espinosa-Lopez2018-hb}. With increasing current densities, the additional heat generated due to the higher overpotential will increase the water flow rates needed for heat management, and could create create implicit constraints on maximum current density levels. At the same time, in the absence of sufficient heat evacuation through increasing feed water flow rate, the cell temperature could rise and and begin vaporizing feed water as well as damage the membrane electrode assembly.

Another operational challenge of dynamic operation is hydrogen crossover to the anode under differential pressure operation, which represents both an energy loss and safety concern. The pressurized cathode provides a driving force for hydrogen to cross the membrane and mix with the oxygen at the anode. At 4\% hydrogen in oxygen, the mixture reaches its lower flammability limit (LFL), presenting serious safety concerns. This crossover is generally greater under differential pressure operation and has been shown to be most problematic at low current densities when there is not enough oxygen production at the anode to keep the hydrogen concentration low \cite{Bernt2020-af}. While the crossover phenomenon has been well documented experimentally, most dynamic techno-economic models of PEM systems exclude any discussion of safety, crossover, or mitigation mechanisms (e.g. recombination catalyst on the anode side) \cite{Christopher_Bryce_Capuano_Morgan_Elizabeth_Pertosos_Nemanja_Danilovic2018-lw,Nicolas_Guillet2014-tq}.

It is common for PEM systems to undergo degradation over time, which requires an increased voltage to the cell for the same current as a result of the increased high frequency resistances in the stack \cite{Hartig-Weis2021-mi}. This degradation is a result of dissolution of catalyst, chemical degradation of the membrane, degradation of the bipolar plate, degradation of the current collector, and manufacturing defects \cite{Feng2017-kz,Weis2019-vj}. The industry standard model for PEM electrolysis techno-economics made by the National Renewable Energy Laboratory and U.S. Department of Energy assumes that the rate of degradation (increase in cell voltage) is independent of operation \cite{Brian_D_James_Daniel_A_DeSantis_Genevieve_Saur2016-wz}. For electrolyzers operating statically (i.e. at constant current density), this constant rate of degradation may hold true. However, experimental studies have shown that degradation rate will be impacted by changes in operating current density \cite{Suermann2019-kz,Rakousky2017-te,Papakonstantinou2020-hj,Alia2019-lw,Li2021-nj,Sun2014-iz,Frensch2019-ff}. To our knowledge, no techno-economic systems of PEMH$_2$ production systems account for this usage-based degradation though it plays a large roll in stack lifetime estimates and therefore replacement costs in these systems. Such operation-dependent degradation may alter the incentives for dynamic operation and consequently impact the levelized cost of hydrogen (LCOH).

Our work seeks to address the above gaps in the techno-economic modeling of PEM electrolyzer systems. Our approach is based on a dynamic optimization framework that co-optimizes the design (stack area and on-site \ch{H2} storage) and operation of the PEM electrolysis-based \ch{H2} production process, with the following unique model elements: a) accounting for dynamic mass and energy balances at 15 min resolution at each electrode as well as flux across the membrane, b) development and use of an empirical correlation for characterizing degradation as a function of current density, which allows us to co-optimize stack replacement times, and c) accounting for safety and temperature related operational constraints and their impact on cost-optimal design and operation of the system. We use the model to investigate the cost-optimal sizing and operation of the PEM system under a range of electricity price scenarios representative of present low-VRE grids and future high-VRE penetration grids, as well as capital cost assumptions for the PEM electrolyzer. Through a systematic scenario-based framework, we isolate the impact of degradation on the levelized cost of hydrogen production. For instance, under 2022 capital cost assumptions for PEM systems, accounting for degradation could increase calculated LCOH by about 45\% due to the use of larger stack sizes (2.3x) to minimize operation at high current densities (and hence degradation) and more frequent stack replacements (every 2 vs. 7 years) also due to degradation. Moreover, accounting for degradation after fixing the design of the system to that obtained without accounting for degradation could increase LCOH by a greater amount of 52\%, which highlights the value of the proposed integrated design and scheduling (IDS) framework. 

In general, scenarios that do not consider degradation tend to have a wider current density distribution,  size larger \ch{H2} storage, and utilize the electrolyzer more than scenarios that account for degradation. Applying the model to future electrolyzer capital cost and electricity price scenarios reveals several interesting findings. First, levelized costs are estimated around \$2.50/kg dominated by cost of electricity and other fixed operating costs. Second, as stack costs decrease, the initially installed stack size increases, which further reduces instances of high current density operation and results in: a) increase time for stack replacement from 2 to 3-5 years depending on the electricity price scenario and CAPEX assumption and b) reduced capacity of installed \ch{H2} storage (0.1 days of \ch{H2} demand vs. 0.5 or greater in the 2022 scenarios). 

Finally, we also investigated the impact of enforcing a safety constraint that limits \ch{H2} build up in the anode to avoid an explosive mixture. We found that without this constraint, the electrolyzer's cost-optimal operation profile tends to favor lower current densities which lowers the overall levelized cost but could result in \ch{H2} concentration in anode exceeding 2\% threshold and 4\% LFL in a few periods.

\section{Methods}

\subsection{Model Overview}
\begin{figure}
     \centering
     \begin{minipage}[c]{\textwidth}
     \includegraphics[width=\textwidth]{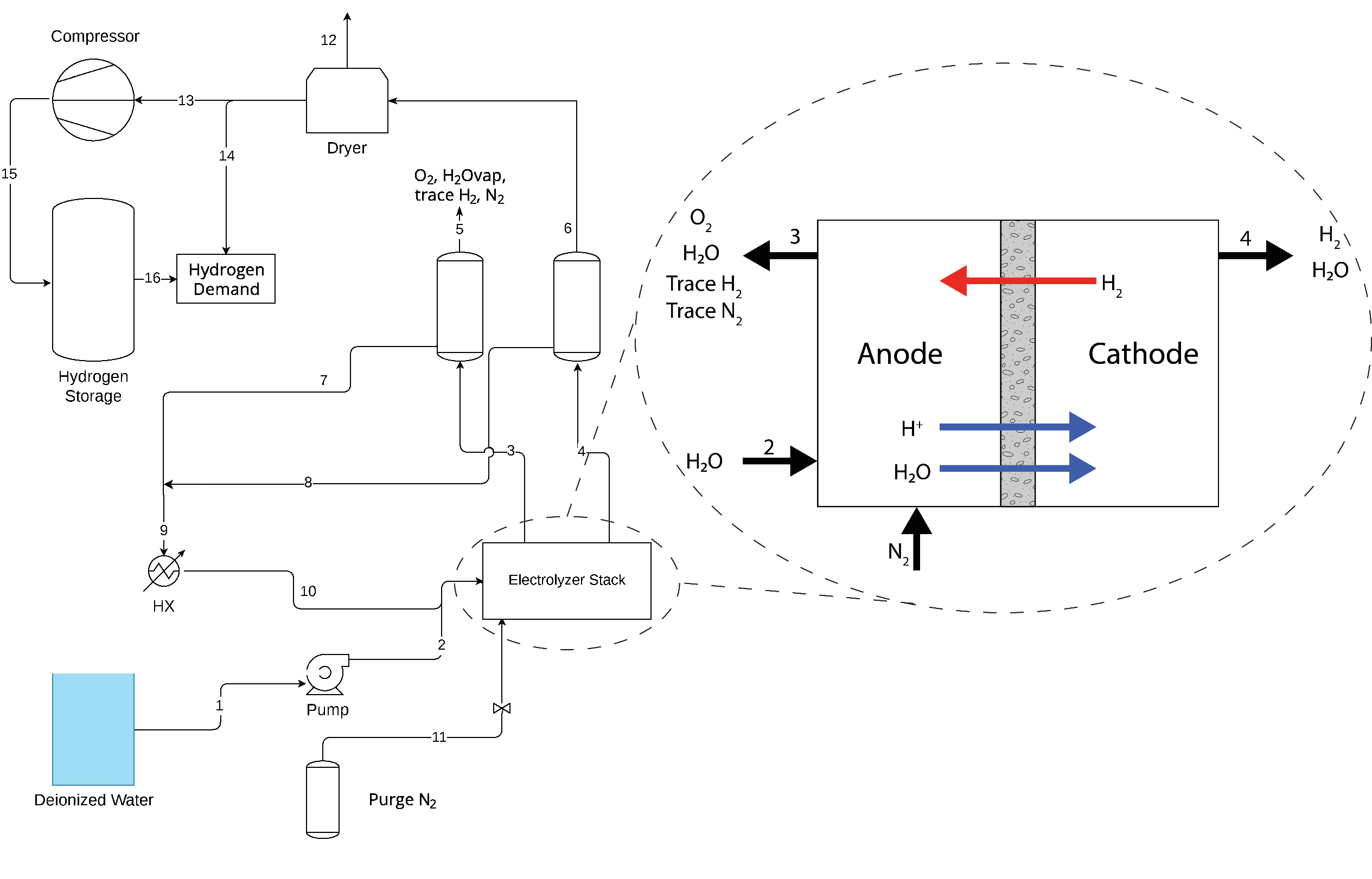} 
     \caption{Flowsheet for the current work. Cell level details with flows across the membrane are shown in the cutout. Downstream treatment of the anode gas stream is not considered. The \ch{H2} stream is dried and can either be compressed and stored to satisfy demand at a later time, or directly used to satisfy demand. Stream 11 is a nitrogen purge stream available for the model to use at a cost as a fail-safe should the concentration of \ch{H2} in \ch{O2} reach the 2\% safety limit.}
     \label{flowsheet}
     \end{minipage}
\end{figure}

In order to explore dynamic operation in the context of heat management, safety, and stack degradation, we develop a 0-D model of the electrolyzer process shown in Figure \ref{flowsheet}, which accounts for electrode-level mass balances, cell-level energy balances as well as electrochemical conversion described by the bottom-up characterization of the polarization curve. Below we provide a brief description of the model, with the complete mathematical description given in the supporting information (SI), along with the model nomenclature in Tables \ref{setnomenclature} - \ref{variablenomenclature}.

Figure \ref{flowsheet} shows the process flow diagram for the electrolyzer and balance of plant which has been adapted from the literature and is consistent with industrial PEM installations \cite{Holst2021-nd,Espinosa-Lopez2018-hb,Caparros_Mancera2020-rv}. Deionized water is pumped into the electrolyzer where the electrolytic reaction takes place. Gases and excess water leave in streams 3 and 4 and are sent to flash drums to separate the vapor from the liquid. It is assumed that a negligible amount of gas is dissolved in the liquid water leaving the flash drum. Water from the flash drum is cooled and recycled back into the electrolyzer stack. The \ch{H2} in stream 6 is dried and then either compressed into \ch{H2} storage or used to satisfy the \ch{H2} demand. Since the focus of the study is on H$_2$ production, the treatment of the anode gas stream, primarily composed of O$_2$, is not considered for this study. Stream 11 is a nitrogen purge stream that is modeled as a backstop to ensure \ch{H2} concentration in the anode stays well below the lower flammability limit. This stream is only used as a fail safe to make sure an explosive mixture is not formed at the anode.

\subsection{General Optimization Formulation}
Given a fixed demand for \ch{H2}, the system as described and the model presented in the SI can be used to minimize the total cost of a PEM electrolyzer system. The general form for such an optimization is shown in Equation \ref{gen_opt_func}.

\begin{equation}
\begin{aligned}
        \min _{x,y} \quad &  PV = C^{stack}(x) + C^{storage}(x) + C^{vOPEX}(x,y) \\
\textrm{s.t.} \quad & f(x,y) =0 \\
& g(x,y) \leq 0 \\
& x \in \mathbb{R}^{n}, y \in \mathbb{R}^m 
\end{aligned}
\label{gen_opt_func}
\end{equation}

The objective function is the present value of the system which is a sum of the cost of the electrolyzer stack, $C^{stack}$, the cost of \ch{H2} storage, $C^{storage}$, and the variable operating costs, $C^{vOPEX}$, which are primarily the cost of electricity and deionized water. The decision variables can be separated into two classes, $x$ which represents the decision variables related to electrolzyer and storage sizing, and $y$, representing the time-dependent operational variables calculated from the relations in $f$ and constraints represented by $g$. A detailed description of the optimization problem is presented in the following sections.

\subsection{Reformulated Model}

Rather than solve a single monolithic non-convex optimization model which is computationally challenging, we apply a decomposition strategy similar to Roh et. al and Chung et. al \cite{Roh2019-vj, Chung2024-vr}. This approach decouples the sizing of the electrolyzer (the outer problem) and the operational optimization given a fixed electrolyzer area and storage capacity (the inner problem) as illustrated in Figure 2 (explained in a later section). 

\subsubsection{Outer Problem}
The outer problem iterates over the total number of cells in the electrolyzer set up, $N_c$, and number of days of \ch{H2} storage $\Gamma_{\ch{H2},stor}$. Each cell is assumed to have an area of 450 \si{\square\centi\meter} represented by $A_{cell}$. The cost of the stack and storage can then be calculated as

\begin{align}
    C^{stack} &= N_{c}A_{cell}p^{stack} + C^{mBoP} + C^{eBoP} \\
    C^{storage} &= \Gamma_{\ch{H2},stor} \dot{m}_{\ch{H2}}p^{storage}
\end{align}

where $C^{mBoP}$ mechanical balance of plant capital cost, $C^{eBoP}$, electrical balance of plant capital cost, $p^{stack}$ is the capital cost of the stack in \$/\si{\square\centi\meter},  $\dot{m}_{\ch{H2}}$ is the daily \ch{H2} production rate in kilograms as calculated by the inner problem, and $p^{storage}$ is the cost of storage in \$/\si{\kilo\gram} \ch{H2}.

\begin{align}
    C^{mBoP} &= \alpha^{mBoP} \dot{m}_{\ch{H2}} \label{mBoP} \\
    C^{eBoP} &= \alpha^{eBoP}P^{max} \label{eBoP}
\end{align}

The mechanical and electrical balance of plant are scaled based on the rate of \ch{H2} production and peak power consumption for the facility, as shown in Equations \ref{mBoP} and \ref{eBoP}. The mechanical balance of plant represents the compressors, pumps, DI water system, and \ch{H2} separation system while the electrical balance of plant includes all electrical components necessary to make the interconnection to the grid including a rectifier. Here, $\alpha^{mBoP}$ is a cost factor in \$/\si{\kilo\gram} \ch{H2}, $\alpha^{eBoP}$ is a cost factor in \$/\si{\kilo\watt}e, and $P^{max}$ is peak power consumption of the system in \si{\kilo\watt}e as calculated by the inner problem. $P_{max}$ is then a parameter passed to the outer problem to calculate the CAPEX for a given outer problem iteration.

\subsubsection{Inner Problem}

The inner problem takes a fixed $N_c$ and $\Gamma_{\ch{H2},stor}$ and minimizes operating cost while meeting the daily production target of $50,000 \si{\kilo\gram\per\day}$. The inner problem can be written as a minimization of the operating cost as defined in Equation \ref{innerproblem}.

\begin{equation}
\begin{aligned}
    \min_{\bf{x}} \quad &C^{vOPEX} = C^{elec} + C^{BoP, \:elec} + C^{water} + C^{\ch{N2}} \\
    \textrm{s.t.} \quad & \textrm{electrochemical model,  Equations \ref{totvolt} -\ref{conduct}}\\
    & \textrm{mass balances, Equations \ref{wateranmol1}-\ref{hydcat2}}\\
    & \textrm{energy balance,  Equations \ref{energybalance} - \ref{energyloss}} \\
    & \textrm{\ch{H2} storage constraints, Equation \ref{deltaequation}-\ref{massrep}}\\
    & \dot{m}_{\ch{H2}} \geq 50,000 \si{\kilo\gram\per\day} \\
    & 60^{\circ}\text{C} \leq T \leq 80^{\circ}\text{C} \\
    & 0.1 \si{\ampere\per\square\centi\meter} \leq i \leq 4 \si{\ampere\per\square\centi\meter} \\
    & y_{\ch{H2},anode} \leq 0.02
\label{innerproblem}
\end{aligned}
\end{equation}

where \textbf{x} represents the primary decision variables, $C^{elec}$ is the annual cost of electricity used by the electrolyzer, $C^{BoP,elec}$ is the annual cost of electricity consumed by the balance of plant, $C^{water}$ is the annual cost of the deionized water, and $C^{\ch{N2}}$ is the annual cost of the nitrogen gas used at the anode for purging.

The model is subject to the mass and energy balances outlined in the SI as well as a demand constraint of 50,000 \si{\kilo\gram\per\day} of \ch{H2}. An operating temperature range of $60^{\circ}$C - $80^{\circ}$C was imposed on the stack. Additionally, a minimum and maximum current density were applied in this study. The minimum current density was chosen based on minimizing degradation at low voltages and the high current density was chosen based on the limit of commercialized PEM today \cite{Weis2019-vj}. In order to ensure 4\% \ch{H2} in \ch{O2} at the anode is never reached a safety factor of 2 was applied making another constraint of a maximum of 2\% \ch{H2} in \ch{O2} at the anode.

Evaluating each variable in the system for 15 minute time periods over the course of a year quickly becomes computationally intractable. To maintain computational tractability, we approximate annual operating costs through modeling operation over representative days at 15 min resolution assuming the hourly price of electricity holds for the entire hour. Time domain reduction using k-means clustering was performed for each electricity price time series data in order to find 7 representative days. Further details of the clustering method can be found in SI \ref{clustering description}. 

The clustering created a mapping $f: d \rightarrow r$. Each real day, $d$, is mapped to a representative day, $r$. To distinguish between variables for real or representative days, any variable with a bar above it is a variable that is indexed by the set of representative days.

Beginning with the cost of electricity for the electrolyzer, 

\begin{align}
    C^{elec} &= \sum_{d=1}^{365} C^{elec}_d\label{cost of elect}\\
    C^{elec}_d &= \int_{0}^{\tau} p^{elec}_{d}(t) i_{d}(t) A \left(V^{undeg}_d(t) + V^{deg,cuml}_d(t)\right) \: dt
\end{align}

where $\tau$ represents the end of a representative day, $p^{elec}$ represents the price of electricity, $i$ is the current density, $A = N_{c}A_{cell}$ is the total electrolyzer area, $V^{undeg}$ is the voltage without considering degradation (based on polarization curve - see Equations \ref{totvolt} -\ref{conduct}), and $V^{deg,cuml}$ is the cumulative increase in voltage due to degradation.

The first term in the integral can be mapped to its respective representative day via $f(d) \rightarrow r$. However, the second term that accounts for the cumulative degradation that changes over the 365 day range and is inter-temporally coupled. So $C^{elec}_d$ becomes:

\begin{equation}
    C^{elec}_d = \int_{0}^{\tau} p^{elec}_{f(d)}(t) i_{f(d)}(t) A \left(V^{undeg}_{f(d)}(t) + V^{deg,cuml}_{d}(t)\right) \: dt
\end{equation}

In accordance with the degradation correlation based on experiential literature values seen in Figure \ref{deg_curve}, we use the following model to represent the change in degradation over a representative day $r$:

\begin{equation}
    \frac{d\bar{V}_{deg,r}}{dt} = \begin{cases}
         a  \label{deg} \quad &\bar{i}_r \leq 1 \\
         a \left[\bar{i}_{r}(t)\right]^2 \quad &\bar{i}_r > 1
    \end{cases}
\end{equation}

Cumulative degradation for real day $d$ can then be written as

\begin{equation}
    V_{d}^{deg,cuml}(t) = V_{d-1}^{deg,cuml}(t) + \delta \bar{V}_{f(d)}^{deg}(t)
\end{equation}

where $\delta \bar{V}_{deg,f(d)}(t)$ comes from solving the differential equation in Equation \ref{deg} up to the time point of interest. Initial degradation at the beginning of the year for a fresh stack is assumed to be 0 \si{\volt}.

Using the mapping $f(d)\rightarrow r$, $C^{elec}_d$ can be written

\begin{equation}
    C^{elec}_d = \int_{0}^{\tau} \bar{p}^{elec}_{r}(t) \bar{i}_{r}(t) A \left(\bar{V}^{undeg}_{r}(t) + V_{d-1}^{deg,cuml}(t) + \delta \bar{V}_{r}^{deg}(t)\right) \: dt
\end{equation}

The annual cost of electricity is then written:

\begin{equation}
    C^{elec} = \sum_{r=1}^7 w_{r} \int_{0}^{\tau} \bar{p}^{elec}_{r}(t) \bar{i}_{r}(t) A \left(\bar{V}^{undeg}_r(t) + \delta \bar{V}_{r}^{deg}\right) \: dt + \sum_{d=1}^{365} \int_{0}^{\tau} \bar{p}^{elec}_{r}(t) \bar{i}_{r}(t) A V_{d-1}^{deg,cuml}(t) \: dt
\end{equation}

where $w_r$ represents the weight of representative day $r$ as determined by the clustering.

$C^{BoP,elec}$, $C^{water}$, and $C^{\ch{N2}}$ can be calculated in a similar manner using the associated representative days and mapping $f$. 

\begin{align}
    C^{BoP,elec} &= \sum_{r=1}^7 w_{r} \int_{0}^{\tau} \left[ \alpha^{mBoP}  \dot{m}_{\ch{H2},r} \bar{p}^{elec}_{r}(t) + W^{compress}\bar{p}^{elec}_{r}(t) \right] dt \\
    C^{water} &= \sum_{r=1}^7 w_{r} \int_{0}^{\tau} \left[ \dot{n}_{\ch{H2O},consum,r} + \dot{n}_{\ch{H2O},vap,5,r} + \dot{n}_{\ch{H2O},vap,6,r}\right]p^{water} dt \\
    C^{\ch{N2}} &= \sum_{r=1}^7 w_{r} \int_{0}^{\tau} \dot{n}_{\ch{N2},11,r} p^{\ch{N2}} dt
\end{align}

where $W^{compress}$ is the work of the \ch{H2} storage compressor, $p^{water}$ is the price of deionized water, and $p^{\ch{N2}}$ is the price of nitrogen. Compressor work is evaluated using a single stage isentropic compression with an isentropic factor of 0.7 as outlined in Kahn et. al and Chung et. al \cite{Chung2024-vr,Kahn2021-jv}.

\subsubsection{H$_2$ Storage}
\ch{H2} storage is necessary to enable meeting base load \ch{H2} demand with flexible electrolyzer operation. We model the state of charge variation in \ch{H2} storage considering both the short-term (intra-day) and long-term (inter-day) changes resulting from hour-to-hour discharging/charging activity.   Following the method of Narayanan et. al \cite{Narayanan2022-uj} and Kotzur et. al \cite{Kotzur2018-rx}, the storage state of charge variation throughout the year is approximated as as a super-position of: a) inter-day state of charge variations modeled over all days of the year and b) intra-day state of charge variations modeled only for the representative days. This approach enables us to respect inter-temporally coupled nature of energy storage operations without the need to model all time periods of the year.

Long duration storage as well as intra-day storage are formulated as follows:

\begin{align}
    \delta_{r} &=  \bar{\Lambda}^{\ch{H2},stor}_r(\tau) - \bar{\Lambda}^{\ch{H2},stor}_r(0) \quad &\forall r \in D_{r} \label{deltaequation}\\
    \Lambda^{\ch{H2},stor}_1(0) &= \Lambda^{\ch{H2},stor}_{365}(\tau) + \delta_{f(365)} \label{wraparound}\\
    \Lambda^{\ch{H2},stor}_{d+1}(0) &= \Lambda^{\ch{H2},stor}_{d}(0) + \delta_{f(d)} \label{realdaycarry}\\
    \Lambda^{\ch{H2},stor}_{d}(0) &= \bar{\Lambda}^{\ch{H2},stor}_{r}(\tau) - \delta_{f(d)} \quad &\forall d \in D \label{realtorep}\\
    \frac{d\Lambda^{\ch{H2},stor}_{d}}{dt} &= \dot{n}_{15,f(d)}(t) - \dot{n}_{16,f(d)}(t) \label{massreal}\\
    \frac{d\bar{\Lambda}^{\ch{H2},stor}_{r}}{dt} &= \dot{n}_{15,r}(t) - \dot{n}_{16,r}(t) \label{massrep} \\
 \Gamma_{\ch{H2},stor} \dot{m}_{\ch{H2}} \label{max_storage} &\geq \max\left(\bar{\Lambda}^{\ch{H2},stor}_r,\Lambda^{\ch{H2},stor}_{d}\right) 
\end{align}

where $\delta_{r}$ is the positive or negative change in storage levels that occurs over representative day $r$ and allows for \ch{H2} storage to carry over between representative periods. Equation \ref{wraparound} represents a wrapping constraint between the last day in the year and the first day in the year. Equation \ref{realdaycarry} allows \ch{H2} storage to carry over between real days. Equation \ref{realtorep} relates the beginning of a real day of storage to the associated beginning of the representative day. Equations \ref{massreal} - \ref{massrep} ensure the mass balance across the \ch{H2} storage is closed over real and representative days. Finally, Equation \ref{max_storage} ensures all storage state of charge variables are less than the maximum amount of \ch{H2} storage. A visual representation of the storage formulation is included in the SI in Figure \ref{storage_form}.

\subsubsection{Implementation Approach}

\begin{figure}
     \centering
     \begin{minipage}[c]{0.7\textwidth}
     \includegraphics[width=\textwidth]{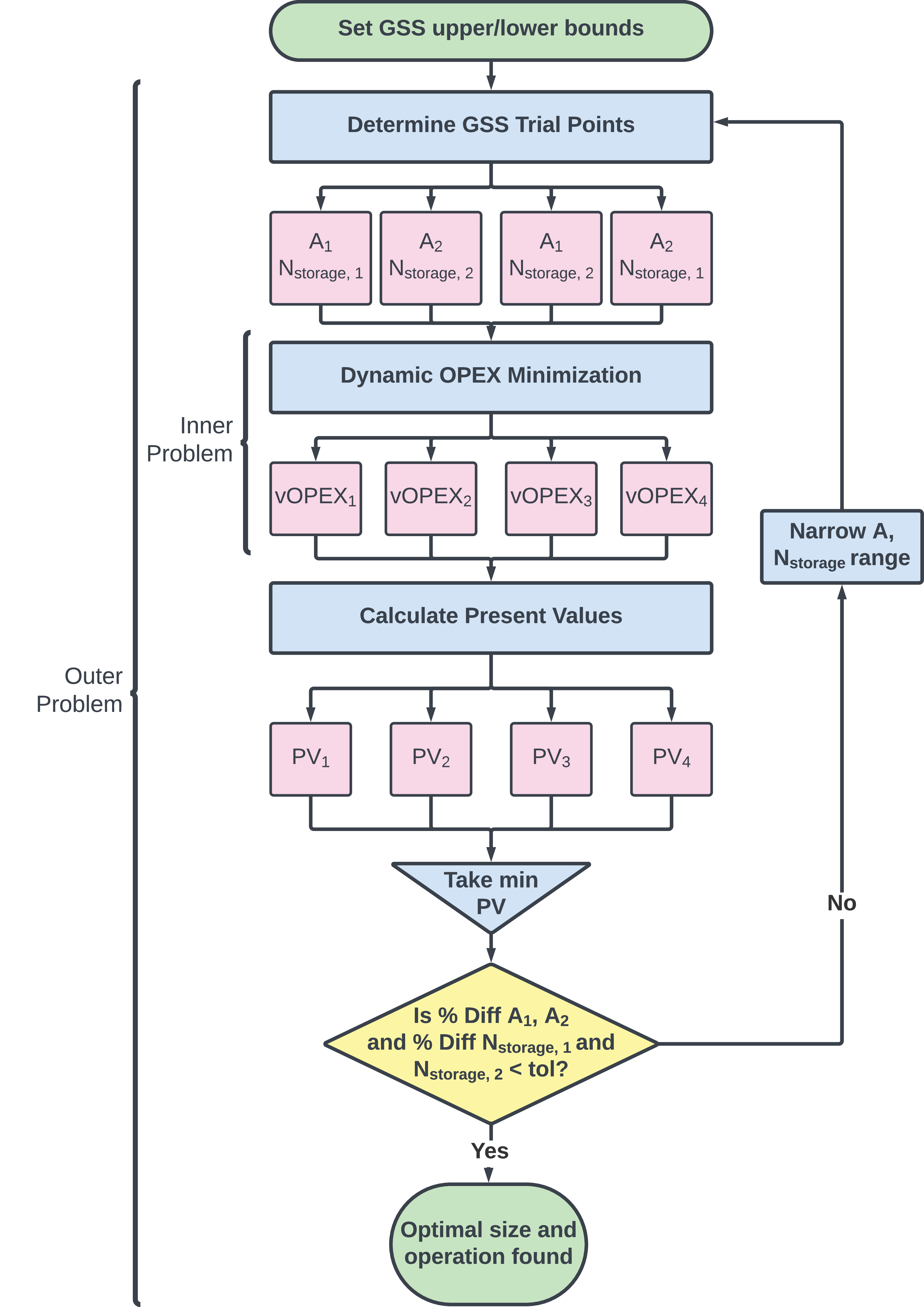} 
     \caption{Summary of the bilevel optimization. The outer problem iterates over number of cells and \ch{H2} storage capacity. The inner problem solves the cost optimal variable operation problem given a fixed number of cells and storage from the outer problem.}
     \label{gss}
     \end{minipage}
\end{figure}

The outer and inner problems can be combined in an iterative manner using the algorithm described in Figure \ref{gss}.

Upper and lower bounds for the electrolyzer area and number of days of \ch{H2} storage are chosen based on physical limits of the electrolyzer in order to satisfy the 50,000 \si{\kilo\gram\per\day} \ch{H2} demand. For this study, an area range of $40,000 - 300,000$ 450 \si{\square\centi\meter} cells and a storage range of $0.1-14$ days were chosen where one day of storage is equivalent to 50,000 \si{\kilo\gram} of \ch{H2}. From the GSS algorithm two trial areas and two trial storages are chosen as is done by Sandhya et al assuming unimodality in each of the search directions \cite{Sandhya-Rani2019-pw}. This creates a total of four trial points in this 2D search space. The operational optimization is run for those 4 trial points which results in four different variable operating costs. For the PEM system, in the limit of very large electrolyzer areas, high capital costs would result in a high present value. In the limit of small areas, the system would need to operate at high current densities which would result in more stack replacement and a high present value. Storage also exhibits a similar unimodal trend at the extremes.

With the size of each system and the variable operating cost estimated, the present value for each of the four trial points can be calculated.  If the trial areas and number of storage days are within the user defined tolerance (0.1\% in this study), then the algorithm terminates and the final solution corresponds to the solution of trial point problems with the lowest present value. The variable operating routine from the associated minimum present value is then the optimal operation routine and the area and number of days of storage from the minimum present value are the optimal sizing of the system. 

If still outside the tolerance, regions in the 2D space where the minimum cannot lie are eliminated. For example, for $A$ and $\Gamma_{\ch{H2},stor}$ where $A_{lb} \leq A \leq A_{ub}$ and $\Gamma_{\ch{H2},stor,lb} \leq \Gamma_{\ch{H2},stor} \leq \Gamma_{\ch{H2},stor, ub}$ and  given four present values that are functions of the two trial areas and two trial storages $PV_A (A_{1},\Gamma_{\ch{H2},stor,1}), PV_B (A_{2},\Gamma_{\ch{H2},stor,1}), PV_C (A_{1},\Gamma_{\ch{H2},stor,2}), PV_D (A_{2},\Gamma_{\ch{H2},stor,2})$, we can eliminate a portion of the search region by finding the minimum of the present values. If $PV_A$ is the smallest present value, we know the minimum will not lie in $A_{2} \leq A \leq A_{ub}$ and $\Gamma_{\ch{H2},stor,2} \leq \Gamma_{\ch{H2},stor} \leq \Gamma_{\ch{H2},stor, ub}$. Thus $A_2$ and $\Gamma_{\ch{H2},stor,2}$ become the new upper bounds. Similar eliminations can be made if the smallest present value is one of the other three values. The algorithm repeats until converged on a locally optimal solution. Due to the non-linearity of the problem, a global optimum is not guaranteed.

The NLP to solve the operational problem was implemented in Python using Pyomo, specifically Pyomo DAE with IPOPT as the non-linear solver \cite{Nicholson2018-mw,Wachter2006-la}. Linear solver MA86 was chosen to use in IPOPT for its ability to handle large problems well \cite{UnknownUnknown-hg}. The bilevel optimization approach with the degradation correlation results in a problem with $\approx 200,000$ continuous variables, $\approx 200,000$ equality constraints, and $\approx 2,500$ inequality constraints. Full details of each run can be found in Table \ref{results} and in the SI in Table \ref{convergence_time}. The problem was run using resources on MIT's SuperCloud HPC \cite{Reuther2018-xn}.

Table \ref{algo_scaling} shows how the bi-level optimization approach scales with the number of representative days. As we increase the number of representative days we expect convergence to a value for storage, number of cells, and LCOH, although this will not be monotonic due to the different effect of each added day to the optimization. Although increasing number of representative days allows for capturing greater day-to-day variability in the underlying electricity price time series, it increases the overall problem size and the computational burden as indicated by the time to convergence and avg. run time per GSS iteration. We choose 7 representative days to strike a balance between computational tractability and accuracy.

\subsection{Data Inputs}
\subsubsection{Techno-economic Data}

For the 2022 case, capital and operating cost assumptions were made consistent with the 2019 version of NREL's H2A model adjusting for inflation from 2019 to 2022 \cite{Brian_D_James_Daniel_A_DeSantis_Genevieve_Saur2016-wz}. The 2030 case capital costs come from Fraunhofer ISE's cost forecast for PEM and alkaline electrolysis \cite{Holst2021-nd}. Summary of the cost assumptions can be found in Table \ref{technoecon}.

\begin{table}
\caption{Techno-economic parameters for model. Consistent with NREL's H2A Model, Chung et. al, and Fraunhofer ISE \cite{Brian_D_James_Daniel_A_DeSantis_Genevieve_Saur2016-wz,Chung2024-vr,Holst2021-nd}. Planned replacement costs are used as the stack replacement cost when a replacement is necessary. }. 
\centering
\begin{tabular}{lcccc}
\toprule
     Techno-economic Parameter                 & 2022          & 2030       & Units  \\ \midrule
Stack CAPEX           & \$2.37        & \$0.79     & \$/cm$^2$                             \\
BoP CAPEX             & 289           & 103        & \$/kWe                             \\
Storage CAPEX         & \$500         & \$300      & \$/kg \ch{H2}                           \\
Site Prep             & \multicolumn{2}{c}{2\%}   & of direct capital\\
Engineering           & \multicolumn{2}{c}{10\%}  &     of direct capital                               \\
Contingency           & \multicolumn{2}{c}{15\%}  &     of direct capital                               \\
Permitting            & \multicolumn{2}{c}{15\%}  &     of direct capital                               \\
Planned Replacement   & \multicolumn{2}{c}{15\%}  &     of direct capital                               \\
Unplanned Replacement & \multicolumn{2}{c}{0.5\%} &     of direct capital                               \\
Overhead              & \multicolumn{2}{c}{20\%}  & of labor cost                      \\
Tax/Insurance         & \multicolumn{2}{c}{2\%}   & of total CAPEX                     \\
BoP Electricity Usage & \multicolumn{2}{c}{5.1}   & kWh/kg \ch{H2}  \\                       \bottomrule
\end{tabular}%
\label{technoecon}
\end{table} 

Using the parameters in Table \ref{technoecon}, capital cost ($C^{CAPEX} = C^{stack} + C^{storage}$), unplanned replacement costs ($C^{UPR}$), planned replacement costs ($C^{PR}$), fixed operating cost ($C^{fOPEX}$), and variable operating cost ($C^{vOPEX}$) can be estimated. Assuming a plant life of 40 years and a discount rate of 8\%, the present value of the project ($PV$) can be estimated as shown in Equation \ref{pv}.

\begin{equation}
    PV = C^{stack} + C^{storage} + \sum_{y=1}^{40}\left[ C^{UPR} +C^{PR} + C^{fOPEX} + C^{vOPEX}\right] \frac{1}{(1+d)^y} \label{pv}
\end{equation}

The direct capital needed to calculate the above costs is the sum of the cost of the stack and cost of the balance of plant. For the scenario run without useage-dependent stack degradation, the replacement rate for the stack was fixed at 7 years \cite{Brian_D_James_Daniel_A_DeSantis_Genevieve_Saur2016-wz}. For the scenarios where degradation is accounted for as shown in Equation \ref{deg}, a degradation rate-informed replacement rate is calculated as follows. Based on the fixed H2A degradation rate, approximately $\Delta^{max}_{deg} \approx 1 \si{\volt}$ of degradation is allowed to occur before the stack needs to be replaced \cite{Brian_D_James_Daniel_A_DeSantis_Genevieve_Saur2016-wz}. The operational optimization of the problem gives the degradation of a fresh stack after one year of operation ($\Delta^{1}_{deg}$). The replacement rate can then be defined as:

\begin{equation}
    r_{stack} = \begin{cases}
         \left\lfloor \frac{\Delta^{max}_{deg}}{\Delta^{1}_{deg}}\right\rfloor \quad &\Delta^{1}_{deg} \leq 1 \\
         1 \quad &\Delta^{1}_{deg} > 1
    \end{cases}
\end{equation}

For implementation purposes, if the degradation of the stack exceeds 1 V in a given year, the stack is allowed to run until the end of the year before being replaced. 

Fixed operating cost is the sum of the cost of labor, overhead, tax and insurance, and material. The labor rate was assumed to be \$70 per hour per worker where 10 workers are needed for a system of this size, and the plant is assumed to operate 350 days per year.

Variable operating cost for the first year is simply obtained from the objective function of the operational optimization. Post-convergence of the operational optimization, the variable operating cost is updated to include degradation for the years leading up to replacement. This assumes that degradation in each future year is the same as the first year after stack replacement.

The amount of \ch{H2} produced in one year ($M_{\ch{H2}}$) can be discounted in a similar way as the present value of the project as seen in Equation \ref{h2pv}. The levelized cost of \ch{H2} is then calculated as the ratio of the two present values.

\begin{equation}
    PV_{\ch{H2}} = \sum_{y=1}^{40} \frac{M_{\ch{H2}}}{(1+d)^y}
    \label{h2pv}
\end{equation}

\begin{equation}
    LCOH = \frac{PV}{PV_{\ch{H2}}}
\end{equation}

\subsubsection{Electrolyzer Performance and Model Assumptions}

\begin{figure}
  \subcaptionbox*{}[0.45\linewidth]{
    \includegraphics[width=\linewidth]{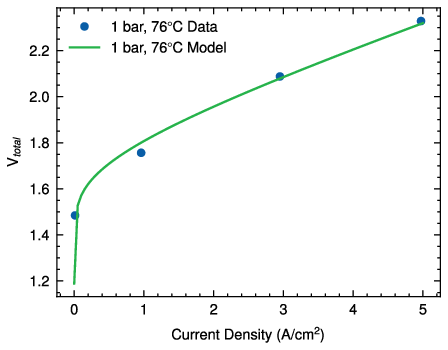}}
  \hfill
    \subcaptionbox*{}[0.45\linewidth]{
    \includegraphics[width=\linewidth]{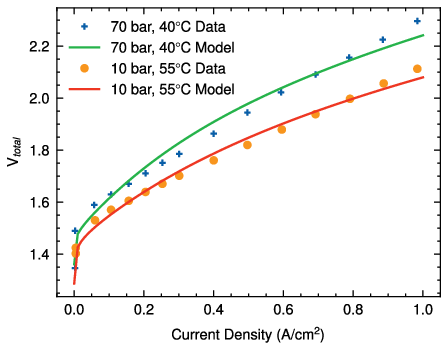}}
    
    \caption{Left Figure: Current voltage relationship at 1 bar and 76$^{\circ}$C. Data from \cite{Lee2020-db}. Fitted parameters $\alpha_{an} = 0.58$, $\alpha_{cat} = 1.28$. Right Figure: Current voltage relationship at high pressure. Data from \cite{Marangio2011-jc}. Fitted parameters at 40$^\circ$C: $\alpha_{an} = 1.9$, $\alpha_{cat} = 0.1$, fitted parameters at 55$^\circ$C: $\alpha_{an} = 1.38$, $\alpha_{cat} = 0.11$ } 
    \label{polarcurves}

\end{figure}

Validation of the electrochemical model without degradation at different temperatures and operating pressures is shown in Figure \ref{polarcurves}. The model fits well across varying current densities, pressures, and temperatures.

Many degradation studies have been done on PEM cells that attempt to characterize the degradation rate of the cell as a function of the applied current density \cite{Suermann2019-kz,Rakousky2017-te,Papakonstantinou2020-hj,Alia2019-lw,Li2021-nj,Frensch2019-ff}. We use data from these studies to construct a usage-based degradation rate. The operating current density and degradation rate have been averaged over the lifetime of the cell and only square and hold wave patterns were taken from the degradation studies \cite{Suermann2019-kz,Rakousky2017-te,Papakonstantinou2020-hj,Alia2019-lw,Li2021-nj,Frensch2019-ff}. Additionally some studies exhibited a negative degradation rate likely to a decrease in ohmic losses from membrane thinning. For this work, those data points were not considered.

\begin{figure}
     \centering
     \begin{minipage}[c]{\textwidth}
     \centering
     \includegraphics[width=0.7\textwidth]{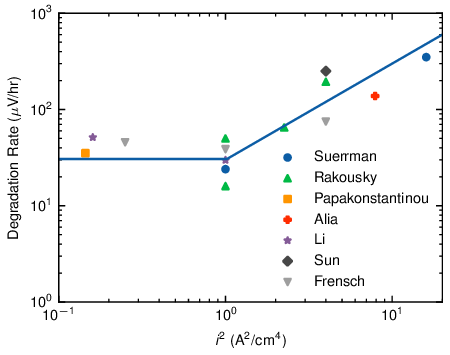} 
     \caption{Degradation rate as a function of the square of the current density plotted on a log-log axis. Data taken from \cite{Suermann2019-kz,Rakousky2017-te,Papakonstantinou2020-hj,Alia2019-lw,Li2021-nj,Frensch2019-ff}. Current density and degradation are time averaged over the operation lifetime. Points shown in the graph here are for square and hold wave patterns. Some cells showed negative degradation rates likely due to membrane thinning; these negative degradation rates were not considered for this study.}
     \label{deg_curve}
     \end{minipage}
\end{figure}

Figure \ref{deg_curve} illustrates the points taken from literature, and Equation \ref{deg_corr} shows the correlation found. The degradation rate increases linearly with the square of the current density for $i > 1 \si{\ampere\per\square\centi\meter}$. The experiments at low current density do not follow this linear trend but instead have a relatively constant degradation rate. The degradation rate was thus modeled as a piece-wise function shown in Equation \ref{deg_corr}. The constant in the equation is a fitted parameter from the data.

Pushing electrolyzers to higher and higher current densities past the 4 \si{\ampere\per\square\centi\meter} maximum presented in this work makes the economics more favorable when degradation is ignored \cite{Chung2024-vr}. Bench scale experimentalists have already begun to push up to 10 \si{\ampere\per\square\centi\meter} \cite{Martin2022-ek}. However, operating at higher and higher current densities could potentially result in increased degradation rates if one assumes the extrapolation of the trend in Figure \ref{deg_curve}. 

\begin{equation}
    \frac{dV_{deg}}{dt} = \begin{cases}
         30  \quad &i \leq 1 \\
         30 \left[i(t)\right]^2 \quad &i > 1 \label{deg_corr}
    \end{cases} 
\end{equation}

\subsubsection{Electricity Price Scenarios}

\begin{figure}
  \subcaptionbox*{}[0.5\linewidth]{
    \includegraphics[width=\linewidth]{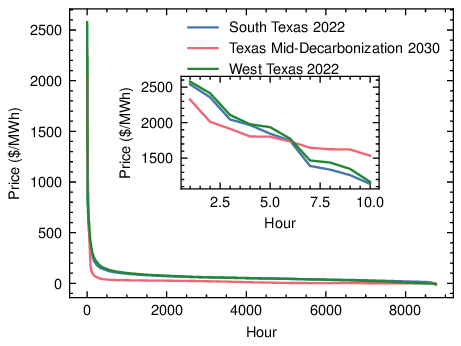}}
  \hfill
    \subcaptionbox*{}[0.5\linewidth]{
    \includegraphics[width=\linewidth]{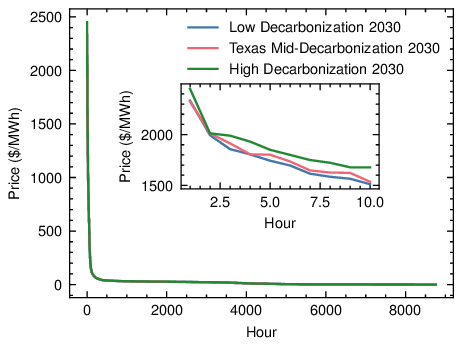}}
    
    \caption{Left Figure: Duration curves for 2022 and 2030 prices used in this study. The south load zone (avg. price: \$62.55) and west load zone (avg. price: \$63.83) prices are for 2022. The 2030 case is the mid-case for NREL's Cambium model for 2030 Texas (avg. price: \$21.43). Right Figure: Duration curves for 2030 prices used in this study. The low decarbonization scenario projects 95\% decarbonization of the grid by 2050 (avg. price: \$21.97). The high decarbonization scenario projects 100\% decarbonization of the grid by 2030 (avg. price: \$22.67).} 
    \label{price_comparison_future}

\end{figure}

Corpus Christi, Texas was used as a case study for this work due to its proximity to existing \ch{H2} demand from the petrochemical sector on the Texas Gulf cost which has prompted commercial interest to deploy GW-scale electrolyzers in that region \cite{Arjona2023-bc}. In order to consider the impact of current and future electricity price scenarios, historical electricity prices from the years 2022 and price projections available for 2030 were used. Electricity prices were taken from 2022 ERCOT South load zone historical pricing \cite{noauthor_undated-so}. For the 2030 scenario, prices were taken from NREL's Cambium future electricity prices model mid-case for Texas \cite{Gagnon_undated-so}. These projected future prices represent a moderately decarbonized grid by 2030. Prices are shown in the left panel of Figure \ref{price_comparison_future}. The average electricity price is higher in 2022 than in 2030. However the 2030 price scenario exhibits more volatile prices.

In order to show the model's sensitivity to 2022 electricity prices, a case with electricity prices from the West Texas load zone in 2022 (another location with many planned electrolyzer projects) was run. For sensitivity to 2030 electricity prices, the Cambium model gives a scenario where the grid is 100\% decarbonized by 2030 (high decarbonization) and 95\% decarbonized by 2050 (low decarbonization). The duration curves for these price series in comparison to the mid-case are shown in the right panel of Figure \ref{price_comparison_future}.

\subsubsection{Limitations of Model}
While this model accounts for many of the aspects of dynamic operation including usage-based degradation, it does have some limitations. In this work we solve a non-linear, non-convex optimization with a local solver IPOPT. There is no guarantee we have converged on the global solution, so a lower cost solution could exist. However, any local minimum will be better than operating naively or statically and a global solution is not strictly necessary in practice. 

Additionally, we model the economics of the system as a price-taker. In other words, we must accept the price of electricity, and our operation has no influence on the prices of electricity. In reality, especially with increasing demand on the grid for electrified industrial processes, dynamic operation to best utilize cheap electricity prices would influence the local cost of electricity, as shown by other studies \cite{Tsay2023-se,He2021-kv}.

\section{Results and Discussion}
In order to demonstrate the sensitivity of the model to capital cost assumptions, electricity price assumptions, and other constraint assumptions, various test cases were run. All scenarios contain the degradation correlation unless ``no use-dependent degradation" is indicated. Capacity factor for the scenarios is calculated by taking the ratio of the energy actually used by the system and the energy used by the system if it were to operate at the maximum current density of 4 \si{\ampere\per\square\centi\meter}. A summary of the tests run with high level results are in Table \ref{results}.
 
\begin{landscape}
\begin{table}
\centering
\caption{High level results of electrolyzer design and scheduling optimization. One day of storage is defined as 50,000 \si{\kilo\gram} of \ch{H2}. Replacement rate is defined as the maximum degradation allowed (1 V) divided by the degradation after 1 year. Annual electrolyzer utilization defined based on maximum current density operation for the 350 working days of the year considered in this model. Convergence time is reported in Table \ref{convergence_time}.}
\resizebox{\columnwidth}{!}{
\begin{tabular}{@{}lccccccccc@{}}
\toprule
Case                       & Price Series     & Total CAPEX (in thousands)          & vOPEX   (in thousands)            & LCOH (\$/kg)    & Days of Storage & Number of Cells (in thousands) & Degradation after 1 Year & Replacement Rate (Years) & Utilization \\ \midrule
Degradation 2022 Case & 2022, South LZ   & \$       366,800 & \$      45,400  & \$         6.60 & 0.51            & 116.2         & 0.45                     & 2.2                      & 25.8\%      \\
No Use-Dependent Degradation             & 2022, South LZ   & \$       203,900 & \$      50,900  & \$         4.56 & 1.39            & 50.1          & --                       & 7                        & 70.1\%      \\
Degradation, West Texas    & 2022, West LZ    & \$       430,800 & \$      46,700  & \$         7.08 & 0.82            & 141.8         & 0.31                     & 3.2                      & 20.5\%      \\
Degradation, Fixed CAPEX   & 2022, South LZ   & \$       239,100 & \$      67,300  & \$         6.92 & 1.39            & 50.1          & 1.97                     & 1.0                      & 67.5\%      \\
High Temperature           & 2022, South LZ   & \$       362,000 & \$      45,600  & \$         6.56 & 0.84            & 110.5         & 0.50                     & 2.0                      & 27.0\%      \\
No Safety Constraint, Fixed CAPEX       & 2022, South LZ   & \$       315,200 & \$      47,500  & \$         6.21 & 0.51           & 116.2        & 0.41                     & 2.4                      & 25.2\%      \\
Degradation 2030 Mid-Case & 2030, Texas Mid  & \$       200,000 & \$        9,400 & \$         2.47 & 0.20            & 156.8         & 0.25                     & 4.0                      & 18.2\%      \\
2030 High CAPEX            & 2030, Texas Mid  & \$       238,300 & \$        9,400 & \$         2.76 & 0.65            & 159.1         & 0.24                     & 4.1                      & 18.1\%      \\
2030 Low CAPEX             & 2030, Texas Mid  & \$       181,900 & \$        9,400 & \$         2.28 & 0.11            & 177.8         & 0.19                     & 5.1                      & 15.8\%      \\
2030 High Decarbonization  & 2030, Texas High & \$       202,300 & \$        9,400 & \$         2.48 & 0.13            & 161.3         & 0.25                     & 4.0                      & 17.8\%      \\
2030 Low Decarbonization   & 2030, Texas Low  & \$       180,400 & \$      11,400  & \$         2.50 & 0.11            & 132.1         & 0.33                     & 3.0                      & 22.1\%      \\ \bottomrule
\end{tabular}
}
\label{results}
\end{table}
\end{landscape}

\subsection*{Degradation vs No Use-Dependent Degradation}

\begin{figure}
     \centering
     \begin{minipage}[c]{\textwidth}
     \includegraphics[width=\textwidth]{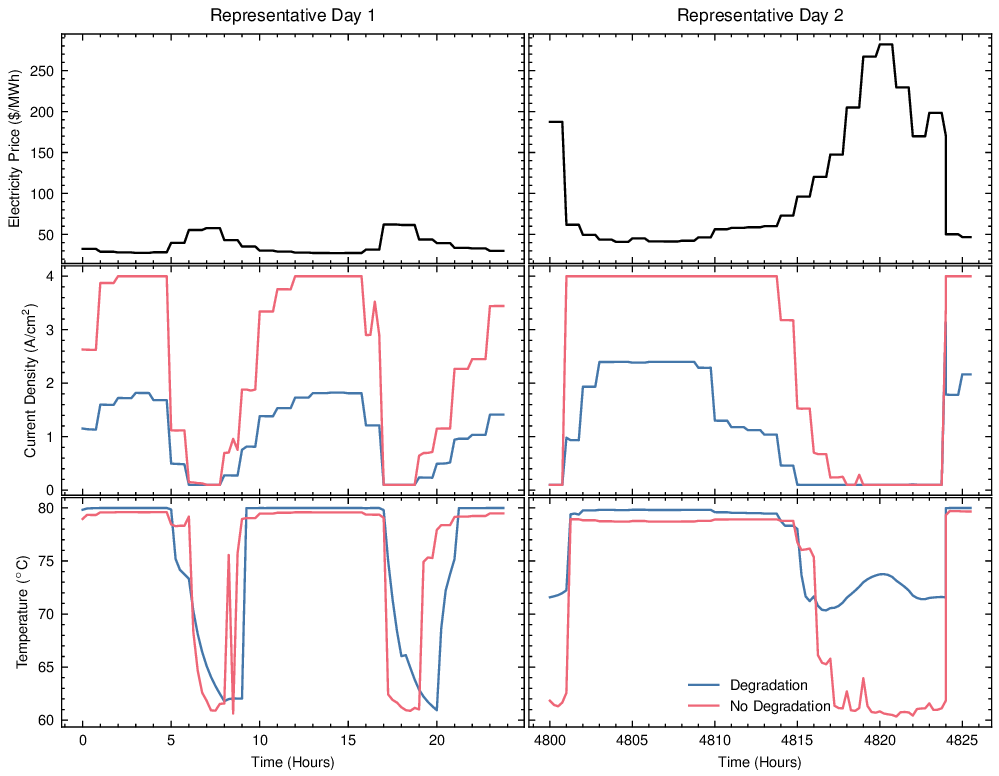} 
     \caption{Impact of modeling use-dependent degradation on optimal electrolyzer operation and temperature profile. Results correspond to 2022 ERCOT South Load Zone price scenario. Two representative days, one from early in the year and one from late in the year, of operation are shown. In general, the case with degradation operates at lower current densities than the case without degradation. During the highest price periods of the day both cases idle at 0.1 \si{\ampere\per\square\centi\meter}.}
     \label{present_results}
     \end{minipage}
\end{figure}

Figure \ref{present_results} highlights the operational impact of incorporating current density dependent degradation in the techno-economic optimization model. Both scenarios produce optimal operating routines that ramp up current density (and therefore production of \ch{H2}) during periods of low prices and switch to idling at 0.1 \si{\ampere\per\square\centi\meter} (the minimum assumed current density) during periods of high electricity prices \cite{Weis2019-vj}. However, because the model with degradation is penalized by accumulating degradation, it operates at much lower current densities in order to minimize the cost of electricity while meeting the demand constraint of 50,000 \si{\kilo\gram\per\day} of \ch{H2}. When electricity prices start to rise, both models switch to the idling state.

\begin{figure}
     \centering
     \begin{minipage}[c]{\textwidth}
     \includegraphics[width=\textwidth]{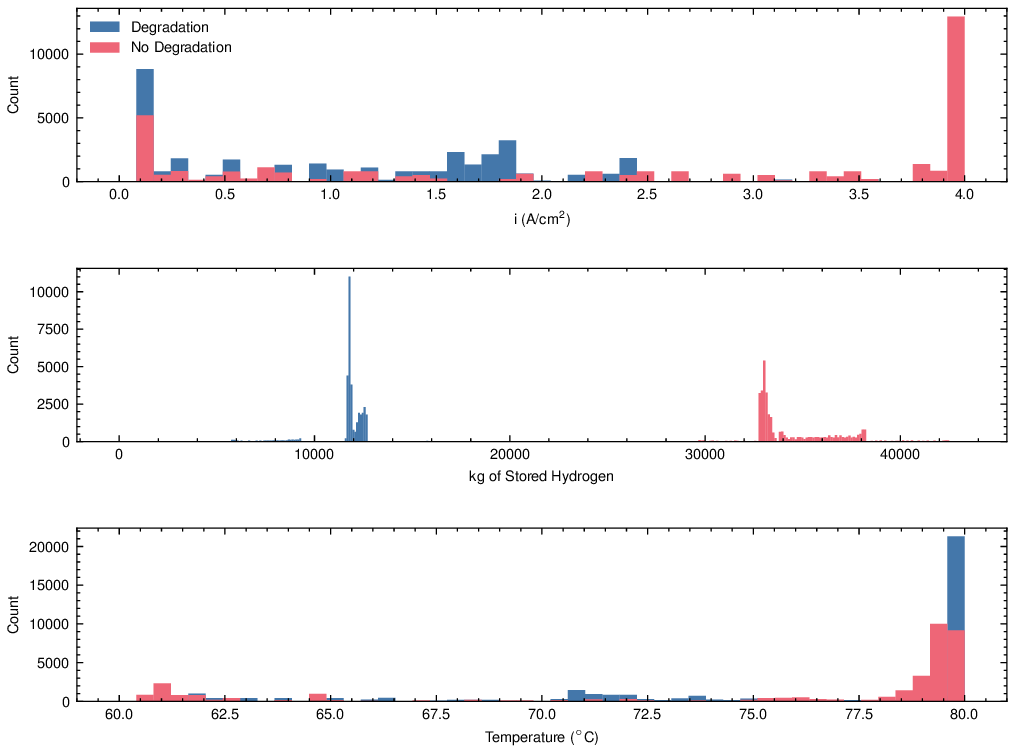} 
     \caption{Cost-optimal current density, storage, and temperature distributions for the cases with and without use-dependent degradation. Accounting for degradation moves the distribution leftwards towards lower average current densities. Results corresponds to model optimization based on 2022 ERCOT South Load Zone scenarios.}
     \label{present_currents}
     \end{minipage}
\end{figure}

The overall effect of degradation on the operation profile can be seen from the current density distribution in Figure \ref{present_currents}. The accumulating degradation causes an increased power demand which forces the model with degradation to favor lower current densities on average, including idling as much as possible while meeting the demand constraint. The net result is that electrolyzer stack utilization (or capacity factor) with degradation is 26\% as compared to 70\% in the case without degradation.

Additionally, due to the increased overpotential from degradation, the case that accounts for degradation operates for more many more hours at the maximum temperature limit than the case without degradation. This is for two reasons: 1) the increased overpotential generates more heat in the system and would result in a higher temperature for the same inlet water flowrate, and 2) operating at a higher temperature lowers the activation and ohmic overpotentials of the cell that partially offset the impact of degradation.

Given today's electricity prices and capital costs for the electrolyzer stack, accounting for degradation results in a 45\% higher LCOH than when use-dependent degradation of the stack is ignored (see Figure \ref{LCOH_final}). The case with no use-dependent degradation has a LCOH that agrees well with the work done by Chung et. al although our model accounts for crossover, static degradation, and the energy balance over the stack \cite{Chung2024-vr}. The case without use-dependent degradation also sized 2.7 times more \ch{H2} storage than the case with it. Intuitively, when use-dependent degradation is not accounted for, the model is incentivized to maximize current density (and production) during periods of low electricity prices and turn down production during periods of high prices. This "bang-bang" operating pattern results in larger need for energy storage. With degradation, there is a reduced incentive for operating at high current densities and thus storage requirements are lower.  Overall use-dependent degradation results in an overpotential of 0.45 V/year that shortens stack replacement to 2.2 years vs. 7 years without modeling use-dependent degradation. A second location in West Texas was also run to show this dynamic operation approach works in more than one electricity market (see Table \ref{results}).

To demonstrate the value of the IDS optimization approach, we ran an operational optimization with use-dependent degradation using the size of stacks and storage from the no use-dependent degradation case. The operational profile can be seen in Figure \ref{fixed_case}. In this highly constrained scenario, the model with degradation is forced to operate at higher current densities much like the case without degradation. Yet the effect of degradation is seen by the model avoiding operation at maximum current density of 4 A/cm$^2$. As a result, the system accumulates 1.97 \si{\volt} of degradation in one year. Consequently, stacks need to be replaced every year at the end of the year. This increases the levelized cost to \$6.92/kg which is a 4.6\% increase over the case with degradation and a 52\% increase over the case without usage-based degradation.

To test the sensitivity of the levelized cost on the degradation correlation and the replacement voltage threshold, two other sensitivity cases were run and compared with the 2022 degradation case: one where the coefficient in Equation \ref{deg_corr} is reduced by half to 15 and one where the maximum allowed replacement threshold voltage is decreased from 1 V to 0.5 V. Results from these runs are summarized in Figure \ref{deg_sensitivity}. Reducing the degradation correlation coefficient results in a 5.2\% decrease in LCOH due to the longer lifetime of these cells (3.04 years), reducing the planned replacement costs. Reducing the replacement threshold voltage results in an 8.5\% increase in LCOH mainly due to more frequent stack replacement (every 2.00 years) due to the lower threshold. Additionally, as seen in Figure \ref{diff_deg_scenarios}, changes in the underlying degradation assumptions do influence operation as well. When the coefficient is lowered to 15, the model can operate at higher current densities than the base case. However, when the voltage threshold is lowered, the model must operate at lower current densities to keep degradation (and therefore replacement rate) low.

\subsection*{Impact of Higher Temperature}

Since the temperature limit was a binding constraint, particularly with use-dependent degradation (Figure \ref{present_results}), we ran a scenario where the upper bound for the temperature of the cell was increased from $80^{\circ}$C to $90^{\circ}$C. The purpose of altering this was to determine if the upper bound of the temperature constraint included in the model was binding. As seen in Figure \ref{high_temp_operation}, operation of the system with the higher temperature bound marginally increases the current density during periods of low electricity prices. This is due to the fact that operating at a higher temperature lowers both activation overpotentials and ohmic overpotentials, which reduces operating costs and incentivizes higher current density operation (see Figure \ref{polarcurves}). For example, at 1 \si{\ampere\per\square\centi\meter} and 60$^\circ$C the voltage is 1.78 \si{\volt} and at 1 \si{\ampere\per\square\centi\meter} and 80$^\circ$C the voltage is 1.7 \si{\volt}.

When looking at the distribution of operating current densities over the year shown in Figure \ref{high_temp_current}, it is clear the higher temperature upper bound case operates at slightly higher current densities generally. These hours of higher current density operation can take advantage of the extra \ch{H2} storage sized in the high temperature upper limit case (See Table \ref{results}). The smaller number of cells in the high temperature upper bound case combined with a greater use of storage results in a \$0.04/kg lower LCOH when compared with the 2022 degradation case (see Figure \ref{LCOH_final}). However, higher current density operation results in more degradation and a more frequent stack replacement rate of 2 years in comparison to the base case replacement rate of 2.2 years (See Table \ref{results}).

\subsection*{Impact of Recombination Catalyst and Safety Constraint}

Industrial PEM membranes come embedded with a gas recombination catalyst on the anode side \cite{Christopher_Bryce_Capuano_Morgan_Elizabeth_Pertosos_Nemanja_Danilovic2018-lw,Nicolas_Guillet2014-tq}. This catalyst lowers the concentration of \ch{H2} at the anode by reacting the \ch{H2} with oxygen to form water. In the original base model, we included a safety constraint. However, in practice, it may be difficult to enforce a safety constraint implying that the recombination catalyst might be the primary mitigation strategy. To evaluate whether the catalyst with 90\% conversion is sufficient to ensure safe operation, we ran the model with the area and storage fixed from the cost-optimal base case with the catalyst present but without a safety constraint. Results are summarized in Figures \ref{no_safety} and \ref{no_safety_hist}.

The distribution of current densities with the relaxed safety constraint shifts to lower current densities. Intuitively, with the constraint relaxed, the model does not need to keep production of \ch{O2} at the anode high and regardless of the rate of crossover can operate in the most cost optimal manner. Relaxation of the safety constraint results in a lowered levelized cost of \$6.21/kg. Though the levelized cost is lower, the \ch{H2} lost to crossover increases from 11,300 kg/year in the base case to 30,400 kg/year in the case without a safety constraint, nearly a 3x change. This \ch{H2} represents lost product as it is not recovered. Replacement rate increases slightly from 2.2 years to 2.4 years due to lower current density operation without the safety constraint, further contributing to the decrease in LCOH.

More important than the change in LCOH is the change in safe operating conditions. Without the safety constraint and with the recombination catalyst, the model exceeds the LFL of 4\% at 1\% of the time points. While only a small percentage of the time operating, crossing into the flammability region of the mixture of gases presents a serious safety concern with dynamic operation. In practice, the composition of the outlet gas of the anode should be monitored continuously. This composition measurement could potentially be integrated into a controller that varies the flowrate of \ch{N2} into the anode to keep \ch{H2} concentrations well below the LFL. 

\subsection*{Future Electricity Prices and Future CAPEX}
To test the sensitivity of these results to future electricity prices and capital costs, we evaluated the model based on a representative 2030 electricity price scenario for Texas and three possible capital cost projections. Capital cost projections in the mid-range case are \$0.79/\si{\square\centi\meter}, consistent with Fraunhofer ISE \cite{Holst2021-nd}. The low capital cost case is \$0.39/\si{\square\centi\meter}, and the high captial cost case is \$1.00/\si{\square\centi\meter}.

\begin{figure}
     \centering
     \begin{minipage}[c]{\textwidth}
     \includegraphics[width=\textwidth]{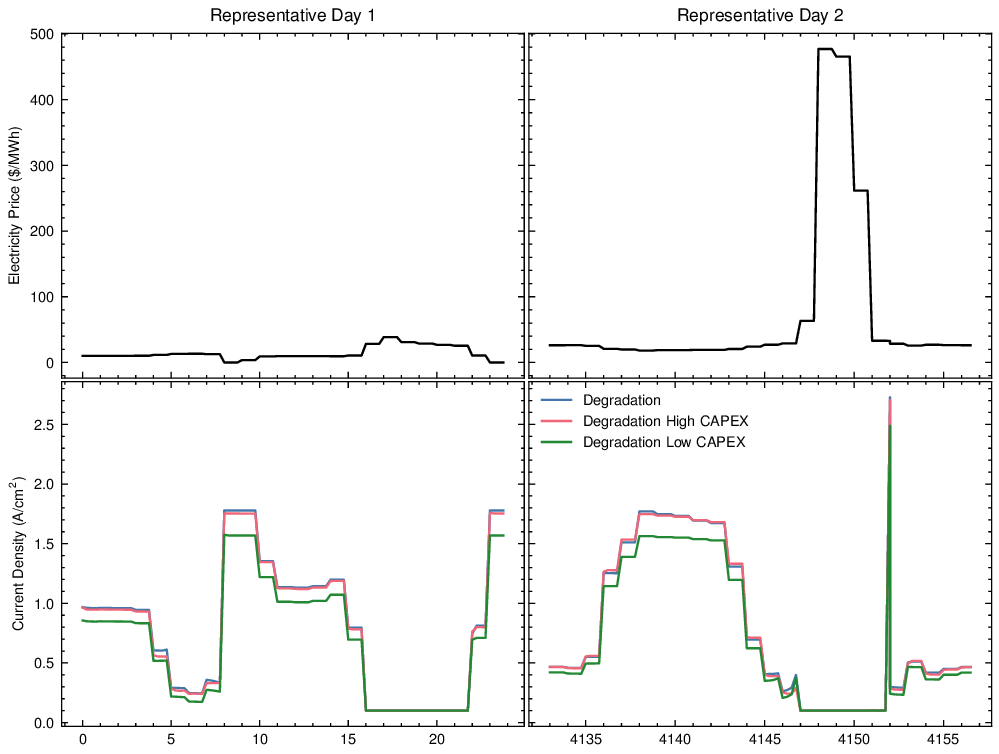} 
     \caption{Two representative days with projected 2030 prices and projected 2030 capital costs. The mid-range CAPEX case has a stack CAPEX of \$ 0.79/cm$^2$, the high CAPEX case is \$1.00/cm$^2$, and the low CAPEX case is \$0.39/cm$^2$. The two cases with the higher capital costs operate quite similarly. The case with the lowest capital cost sizes many more cells and is able to therefore operate at lower current densities. Temperature profile included in Figure \ref{Temp_Only}.}
     \label{future_operation}
     \end{minipage}
\end{figure}

\begin{figure}
     \centering
     \includegraphics[width=\textwidth]{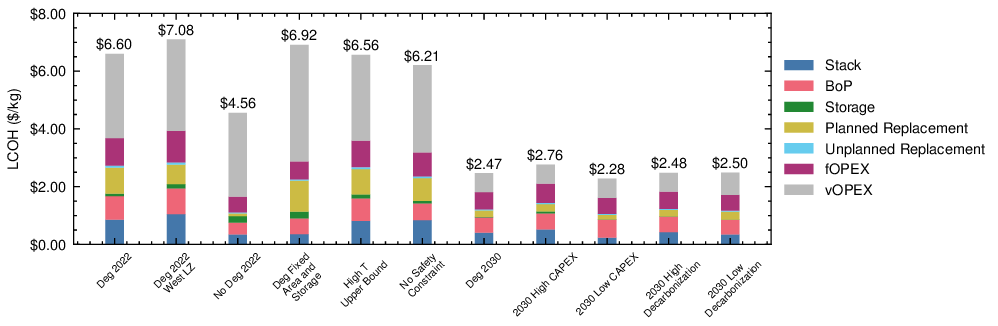} 
     \caption{Levelized cost of \ch{H2} for each of the cases. CAPEX Total is the capital cost of the stacks and downstream equipment, Planned Replacement is the cost for planned replacements of the stack as they age, Unplanned Replacement is the unplanned replacement cost as defined in Table \ref{technoecon}, fOPEX is the fixed operating cost, and vOPEX is the variable operating cost which is the cost of electricity and the cost of water used to feed and cool the stacks. Scenarios reported here are the same as reported in Table \ref{results}.}
     \label{LCOH_final}
\end{figure}

\begin{figure}[!h]
    \centering
     \includegraphics[width=\textwidth]{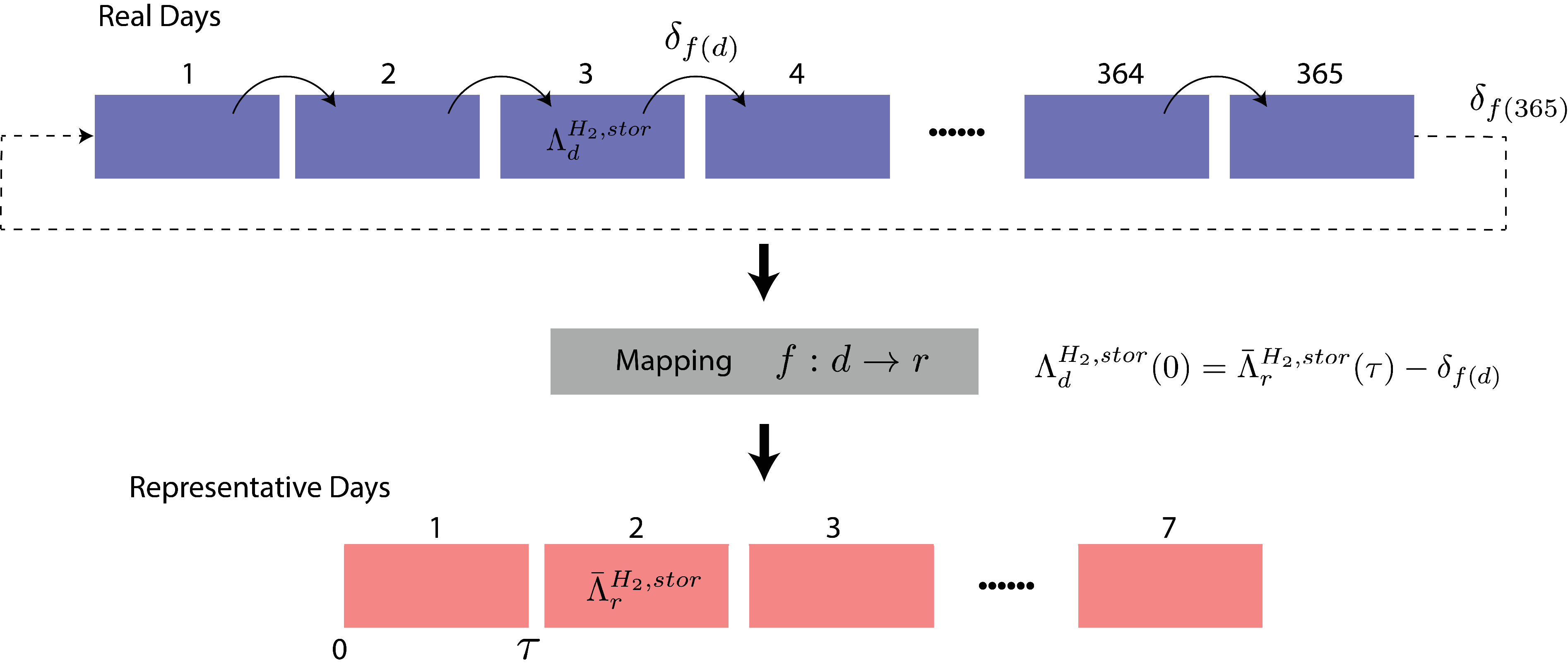} 
     \caption{Storage formulation and connection between representative days and real days. Each real day has an hour to hour changing amount of storage $\Lambda^{H_2,stor}_d$. Each real day is mapped to a representative day by the k-means clustering mapping $f: d\rightarrow r$. Thus each representative day has an hour to hour changing amount of storage $\bar{\Lambda}^{H_2,stor}_r$. $\delta_r$ allows for carry over of storage between real days. To relate the representative days to real days, the beginning of a real day of storage must match the end of the corresponding representative day minus the carry over associated with the representative day which is summarized in the equation in this figure.}
     \label{storage_form}
\end{figure}

\begin{figure}[!h]
     \centering
     \begin{minipage}[c]{\textwidth}
     \includegraphics[width=\textwidth]{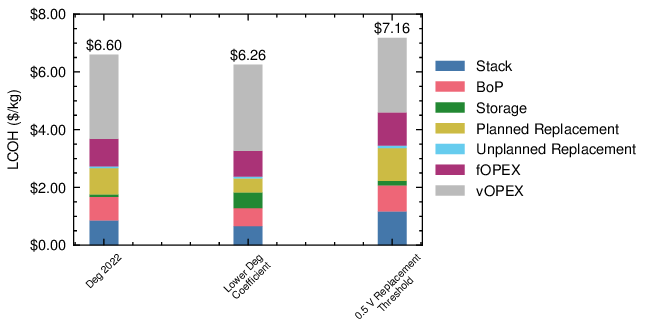} 
     \caption{Sensitivity to the degradation correlation coefficient being reduced to 15 (middle bar) and the replacement threshold voltage being reduced to 0.5 V (right bar) are compared against the base degradation case (left bar). Somewhat unsurprisingly a less aggressive degradation rate results in a lower levelized cost. Reducing the maximum amount of degradation that can occur before stack replacement increases LCOH.}
     \label{deg_sensitivity}
     \end{minipage}
\end{figure}

\begin{figure}[!h]
     \centering
     \begin{minipage}[c]{\textwidth}
     \includegraphics[width=\textwidth]{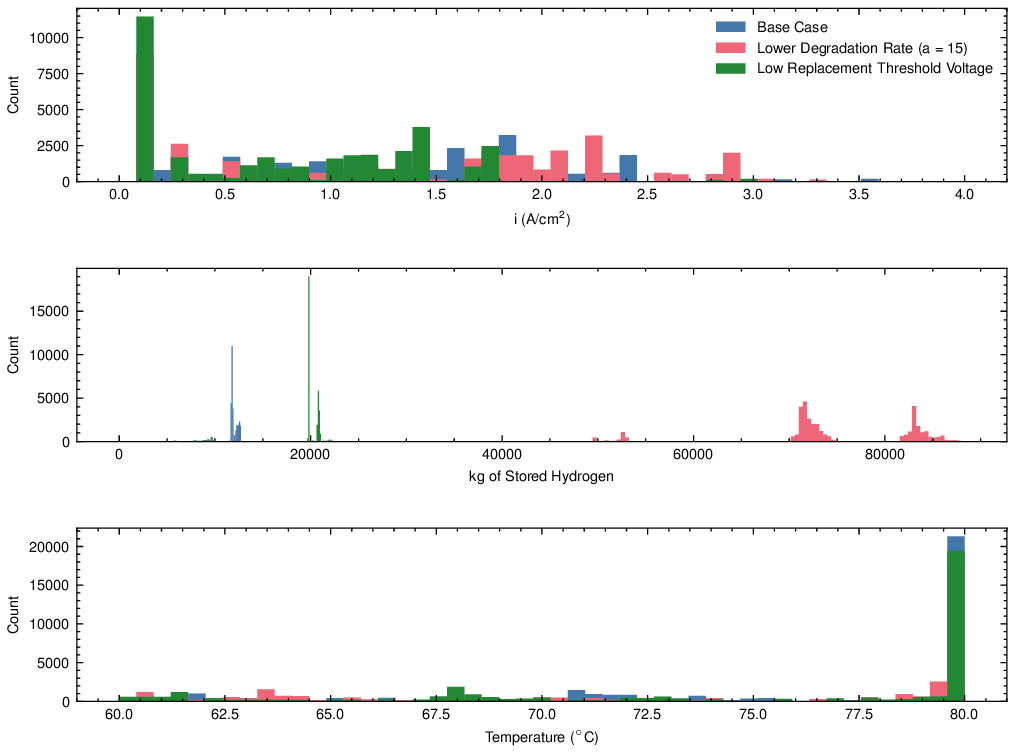} 
     \caption{Changing the underlying assumptions on degradation shifts the operating profile from the base case. With less degradation occuring when a = 15, the model is free to operate at higher current densities than the base case. Additionally when the replacement threshold voltage is lowered the model is incentivized to keep replacement rates as high as possible by degrading as little as possible.}
     \label{diff_deg_scenarios}
     \end{minipage}
\end{figure}

\begin{figure}[!h]
     \centering
     \begin{minipage}[c]{\textwidth}
     \includegraphics[width=\textwidth]{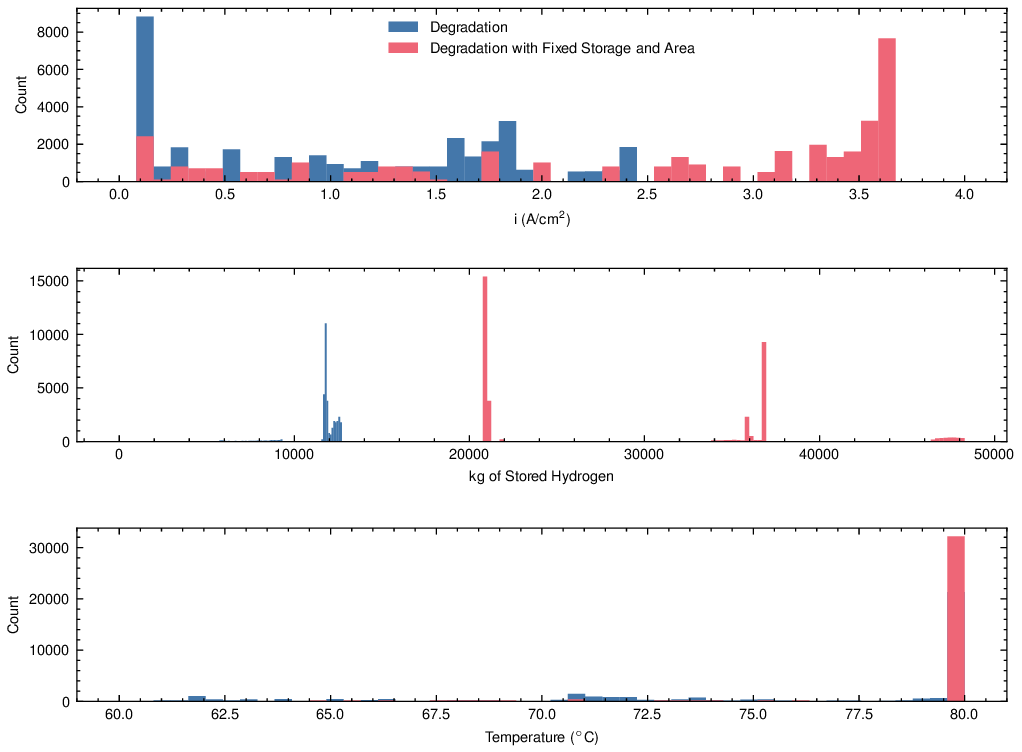} 
     \caption{Current density, storage, and temperature distributions for the base degradation case and the degradation case run with the stack size and storage size of the no degradation case. Fixing the stack size and storage results in having to operate at much higher current densities and more frequent replacements.}
     \label{fixed_case}
     \end{minipage}
\end{figure}

\begin{figure}[!h]
     \centering
     \begin{minipage}[c]{\textwidth}
     \includegraphics[width=\textwidth]{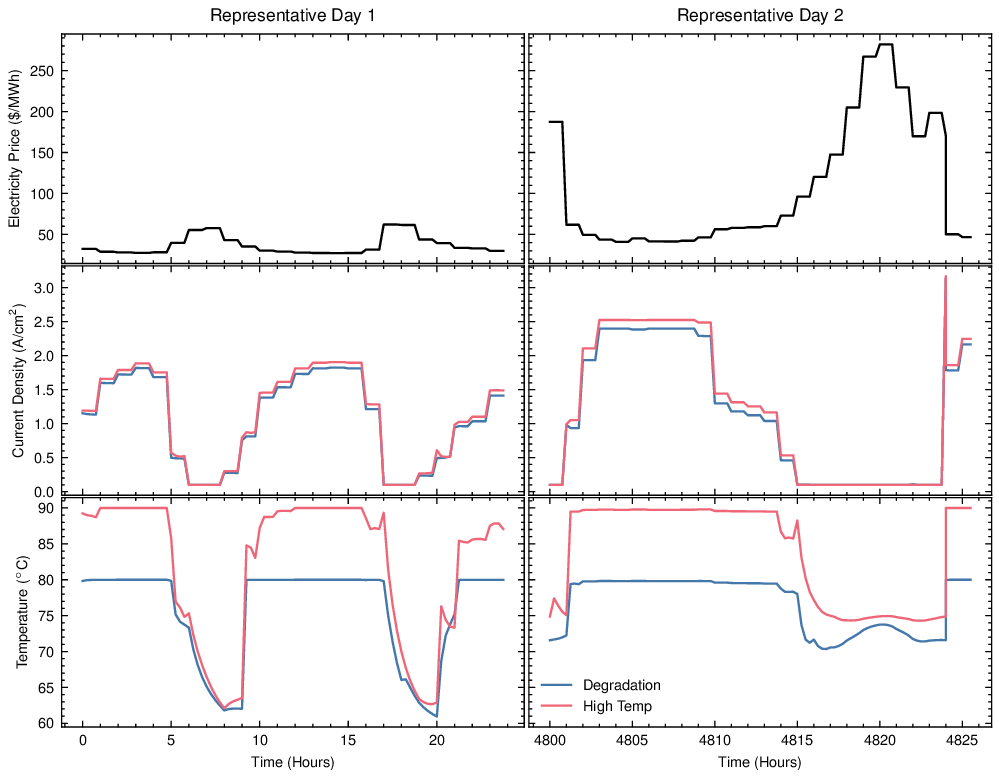} 
     \caption{Current density profile and temperature profile of scenario where the upper bound of the temperature limit is changed from 80$^\circ$C to 90$^\circ$C. Current density operation is nearly the same for both cases with the case with a higher temperature upper bound operating at slightly higher current densities. This coincides with a larger amount of storage sized for this system.}
     \label{high_temp_operation}
     \end{minipage}
\end{figure}

\begin{figure}[!h]
     \centering
     \begin{minipage}[c]{\textwidth}
     \includegraphics[width=\textwidth]{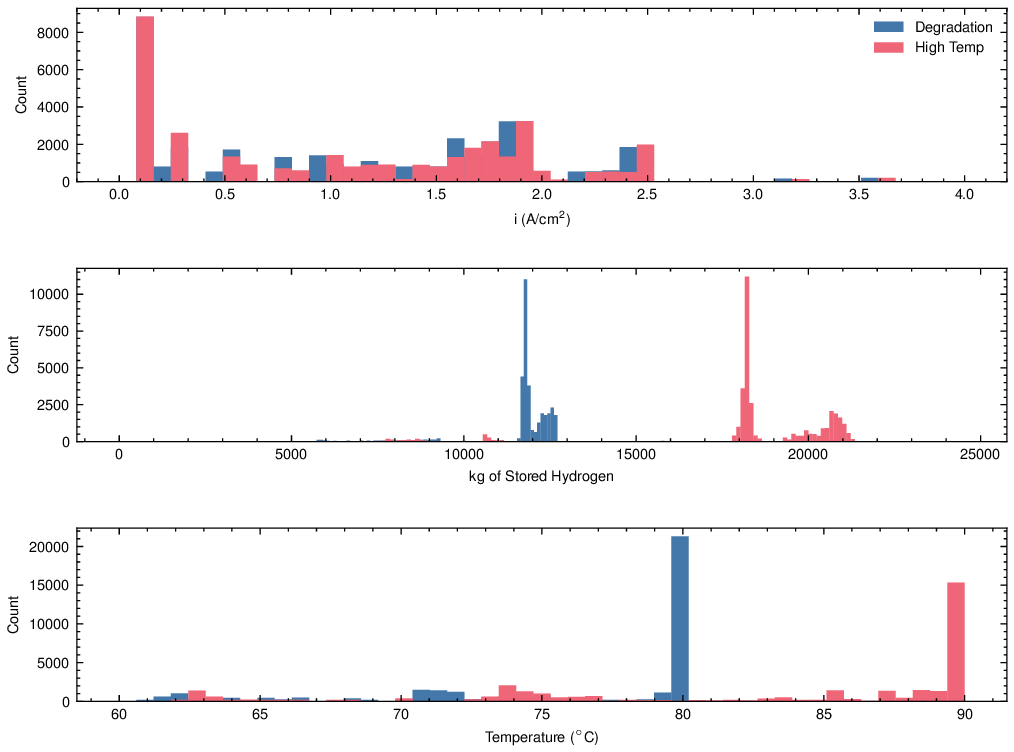} 
     \caption{Higher temperature upper bound case current density, storage, and temperature distribution. The high temperature case operates at slightly higher current densities overall and as a result sizes and utilizes a greater amount of storage than the base degradation case.}
     \label{high_temp_current}
     \end{minipage}
\end{figure}

\begin{figure}[!h]
     \centering
     \begin{minipage}[c]{\textwidth}
     \includegraphics[width=\textwidth]{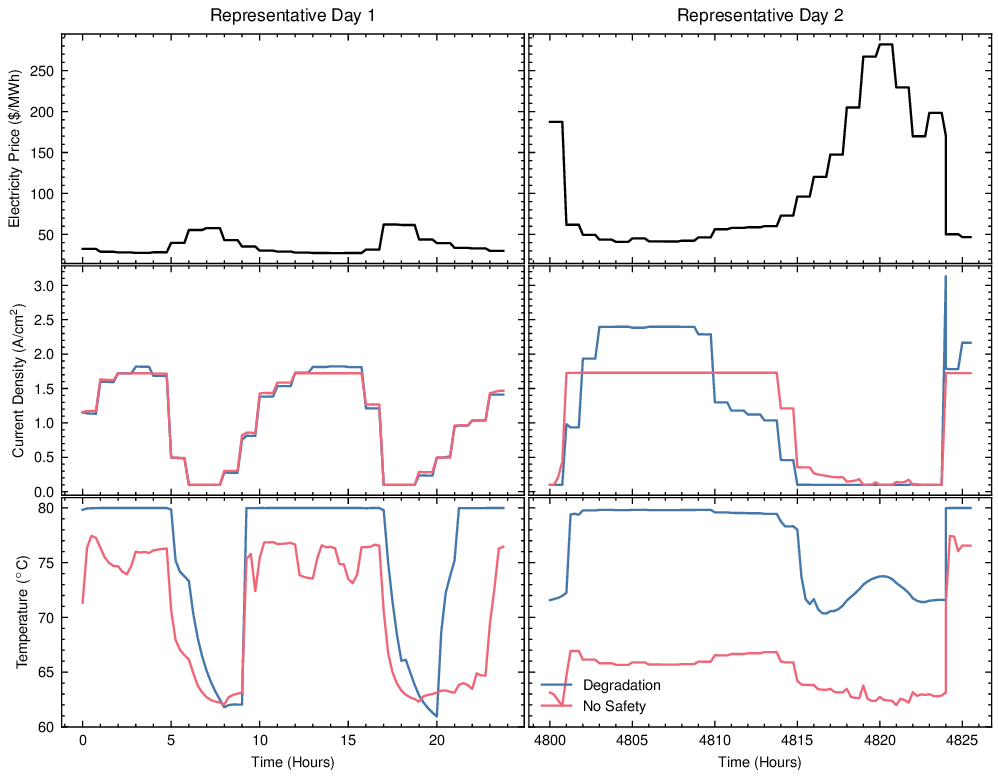} 
     \caption{Operational profiles for the case with no safety constraint. The case without a safety constraint operates at lower current densities since it does not have maintain O2 production at the anode to keep the H2 concentration below lower flammability limit.}
     \label{no_safety}
     \end{minipage}
\end{figure}

\begin{figure}[!h]
     \centering
     \begin{minipage}[c]{\textwidth}
     \includegraphics[width=\textwidth]{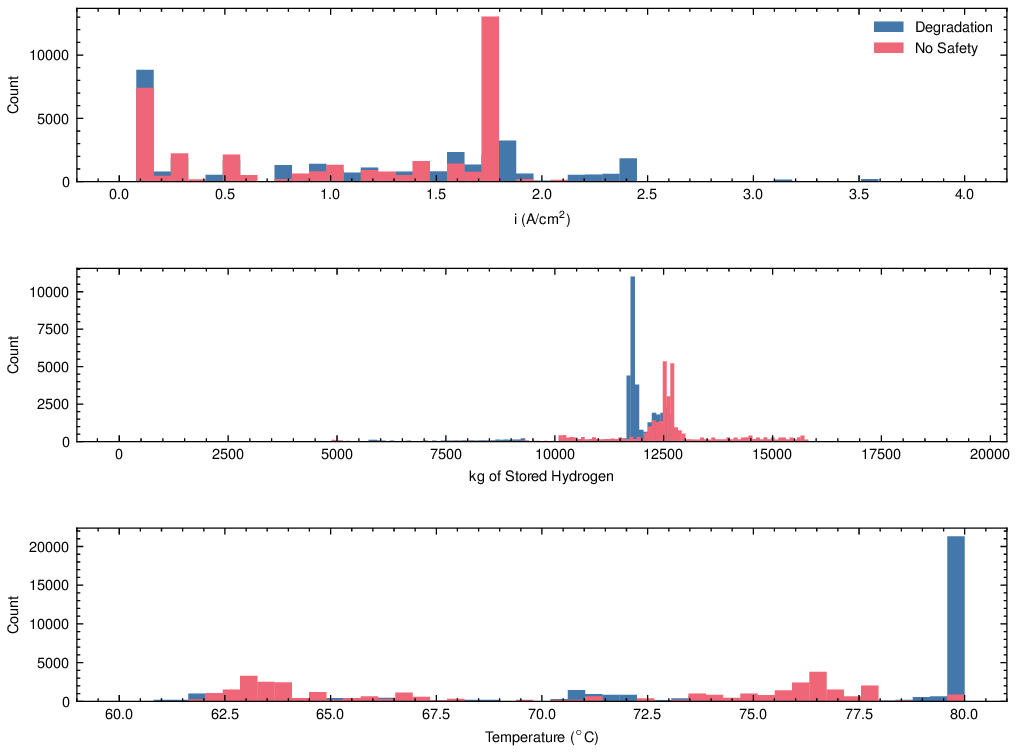} 
     \caption{Current density, hydrogen storage, and temperature distributions for the case run without a safety constraint and the base degradation case. With no incentive to keep hydrogen concentration at the anode low, the current density distribution shifts to lower current densities.}
     \label{no_safety_hist}
     \end{minipage}
\end{figure}

The operational profile in Figure \ref{future_operation}, shows that the 2030 mid-range CAPEX and high CAPEX case operate at similar current densities due to their similarly sized stacks and storage (see Table \ref{results}). Due to the increased cost of stacks in the high CAPEX case, the LCOH is 12\% more than the mid-case (see Figure \ref{LCOH_final}). For the lowest capital cost scenario, the number of cells increases by 13\% and the amount of storage decreases by 47\% compared to the mid-range case. With the greater number of cells, the scenario with the lowest capital cost can operate at lower current densities to avoid high degradation rates and keep operating costs lower. This results in a 7.7\% decrease in LCOH from the mid-range case.

In order to test sensitivity to future electricity prices, another two scenarios with differing electricity prices were run with the mid-range capital costs scenario as shown in Figure \ref{price_comparison_future}. Figure \ref{decarb} shows the differences in operation for the three different electricity price scenarios: the mid-decarbonization case, low decarbonization, and high decarbonization. While we do see the lowest decarbonization case able to operate at higher current densities, and a much larger sized storage for the high decarbonization case, ultimately, the LCOH is quite similar between scenarios indicating a weak relationship between LCOH and electricity price for this particular location (see Figure \ref{LCOH_final}).

\section{Conclusion}
We developed a techno-economic optimization model for integrated design and scheduling of  a PEM electrolyzer that respects temperature constraints, safety constraints, and models degradation of the electrolyzer stack as a function of its use. The model relies on an empirical correlation linking current density with degradation rate derived from experimental data in the literature, and is solved to co-optimize electrolyzer stack, balance of plant and storage capacity along with their operation. To enable computational tractability we employ number of techniques including approximating annual operations via operations over representative days, as well as 2-D GSS search algorithm that decomposes the investment and scheduling operating facilitates identification of locally optimal solutions of the resulting optimization model. We applied the model to explore the techno-economics of such a system and calculate the LOCH for various operating scenarios including current and future prices of electricity and stack capital costs.

Dynamic operation overall results in binding constraints on both temperature and safety limits. Heat management and safety of PEM systems must continue to be an area of focus for the operators of PEM systems if dynamic operation is to be successfully implemented. The monitoring and manipulation of stack temperature and anode gas composition will require detailed process control.

Using the correlation outlined in this work, accounting for degradation resulted in a $\approx$2 year stack lifetime in comparison to the 7 year assumed lifetime in the 2022 scenario. The increase in stack replacement frequency could potentially increase the materials burden on critical minerals like Ir, unless recycling becomes commonplace. Previous works have highlighted the scarcity of Ir especially when considering aggressive decarbonization scenarios \cite{Minke2021-io,Riedmayer2023-nw}. The increased replacement frequency associated with dynamic operation could exacerbate this issue even with replacement rates projected to increase to $\approx 4$ years by 2030. While some experimental work has begun to address this issue, significant advancements in PEM electrolyzer efficiencies, a decrease in Ir loadings in PEM, and a focus on PEM catalyst recycling are critical going forward \cite{Bernt2018-rj}.

The LCOH is higher in the 2022 case with degradation when compared to running without degradation. This indicates an underestimation of LCOH when operating dynamically in models that do not account for use-dependent degradation. We also show that optimizing design without accounting for use-dependent degradation and subsequently incurring these degradation costs leads to 4.6\% higher LCOH than co-optimized design and operations case. Additionally, not accounting for stack degradation leads to an over-sizing in hydrogen storage.

Dynamic operation does indeed make more cost effective hydrogen under present cost scenarios, and we show that dynamic operation makes even more sense under future electricity and capital cost scenarios by significantly minimizing the variable operating cost of electricity (see Figure \ref{LCOH_final}).

While the results presented in this paper highlight issues with cost optimal design and operation of PEM electrolyzers, there are some limitations to this analysis. Our work sought to capture degradation behavior of PEM stacks, however, the degradation correlation was found using the limited data available in literature. Standard procedures for accelerated degradation of cells are needed in order to gather reliable data that then can be used in models. More in depth degradation studies that not only look at the effect of operating current density but also temperature and catalyst loading effects are needed. One other weakness of our approach is the fact that we solve a non-convex NLP with no guarantee of finding a global minimum. Indeed for a problem of this size, it could be expected that there would be multiple minima in the design and operation space.

Most importantly, the framework presented here to evaluate PEM electrolysis and hydrogen production can be applied to the design of other electricity-intensive processes. With growing interest in electrochemical/electricity driven processes, model set up and system design can be readily adapted to study other processes and systems. This framework can help quickly and effectively evaluate different processes on the basis of their economics while modeling the underlying physical phenomenon and dynamics.

\section*{Acknowledgements}
The authors would like to acknowledge Edward Graham for his insightful contributions to this work. Funding for this work was provided by Analog Devices Inc.

\section*{Data Availability Statement}
The model and data used to produce all figures and results will be provided upon reasonable request.

\newpage 
\printbibliography
\end{refsection}

\clearpage
\section{Supporting Information}
\begin{refsection}

\beginsupplement

\subsection{Mass and Energy Balance and Electrochemical Relations}

The following is a detailed description of the mass and energy balances as well as the electrochemical relations used in this work. Tables \ref{setnomenclature} - \ref{variablenomenclature} outline the nomenclature used in this work.

\begin{table}
\caption{Set Nomenclature}

\centering
\begin{tabular}{llccc}
\toprule
Name & Description & Units & Value & Notes\\ \midrule
Set Indices \\ \hline
$i \in I$ & Chemical species & - & \{\ch{H2},\ch{O2},\ch{H2O},\ch{N2}\} & - \\ 
$j \in J$ & Stream numbers & - &\{1,2,...,16\} & -\\
$k \in K$ & Anode/Cathode & - &\{\text{an},\text{cat}\} & -\\
$r \in D_r$ & Representative days & - &\{1,2,...,7\} & -\\
$d \in D$ & Actual days & - & \{1,2,...,365\} & -\\

\bottomrule
\end{tabular}

\label{setnomenclature}
\end{table}

\begin{table}
\caption{Parameter Nomenclature}
\resizebox{\columnwidth}{!}{%
\centering
\begin{tabular}{llccc}
\toprule
Name & Description & Units & Value & Notes\\ \midrule

 $F$ & Faraday's constant & \si[per-mode=fraction]{\coulomb\per\mol}  & 96,485 & -\\

 $A_{cell}$ & Catalytic active area & \si{\square\centi\meter} & 450 & \cite{Brian_D_James_Daniel_A_DeSantis_Genevieve_Saur2016-wz} \\

 $A_{mem}$ & Area of membrane/Size of one cell & \si{\square\centi\meter} & 450 & \cite{Brian_D_James_Daniel_A_DeSantis_Genevieve_Saur2016-wz} \\

 $\delta_{el}$ & Thickness of electrode & \si{\centi\meter} & $8.0 \times 10^{-3}$ & \cite{Abdin2015-nl} \\

  $v_{an}$ & Anode volume & \si[per-mode=fraction]{\mL} & $\approx \delta_{el}A_{mem}$ & - \\

   $v_{cat}$ & Cathode volume & \si[per-mode=fraction]{\mL} & $\approx \delta_{el}A_{mem}$ & - \\

 $\delta_{mem}$ & Membrane thickness & \si[per-mode=fraction]{\centi\meter} & $1.75 \times 10^{-2}$ & \cite{Liso2018-iy} \\

 $n_{g}$ & Electro-osmotic drift coefficient& unitless & - & \cite{Medina2010-lh}, Equation \ref{waterng} \\

 $\Delta G^{\circ}$ & Standard free energy of reaction & \si[per-mode=fraction]{\kilo\joule\per\mol} & $237.2$ & \cite{Liso2018-iy} \\

  $R_{gas}$ & Gas constant & \si[per-mode=fraction]{\joule\per\mol\per\K} & $8.314$ & - \\

$T_{ref}$ & Reference temperature for thermo & \si[per-mode=fraction]{\K} & $298$ & \cite{Liso2018-iy} \\

$H_{i}$ & Enthalpy of species $i$ & \si[per-mode=fraction]{\joule\per\mole} & - & \cite{BurgessUnknown-hi} \\

$C_{p,th}$ & Lumped thermal capacitance & \si[per-mode=fraction]{\joule\per\K} & - & Scaled from \cite{Espinosa-Lopez2018-hb} \\

$R_{th}$ & Thermal resistance of stack & \si[per-mode=fraction]{\second\per\joule\per\K} & - & Scaled from \cite{Espinosa-Lopez2018-hb} \\

$i_{0,an,ref}$ & Exchange current density at $T_{ref}$ & \si[per-mode=fraction]{\ampere\per\square\centi\meter} & $5 \times 10^{-12}$ & \cite{Liso2018-iy} \\

$i_{0,cat,ref}$ & Exchange current density at $T_{ref}$ & \si[per-mode=fraction]{\ampere\per\square\centi\meter} & $1 \times 10^{-3}$ & \cite{Liso2018-iy} \\

$\alpha_{an}$ & Bulter-Volmer equation parameter & - & - & Fitted parameter \\

$\alpha_{cat}$ & Bulter-Volmer equation parameter & - & - & Fitted parameter \\

$\gamma_{M,an}$ & Anode roughness factor & - & 1198 & Equation \ref{roughness}\\

$\gamma_{M,cat}$ & Cathode roughness factor & - & 286 & Equation \ref{roughness}\\

$\varphi_{I}$ & Fraction of metal catalyst in contact with ionomer & - & 0.75 & \cite{Liso2018-iy}\\

$m_{M,an}$ & Catalyst loading at anode& \si[per-mode=fraction]{\gram\per\square\centi\meter} & $0.9\times10^{-3}$ & \cite{Bernt2020-ne}\\

$m_{M,cat}$ & Catalyst loading at cathode & \si[per-mode=fraction]{\gram\per\square\centi\meter} & $0.3\times10^{-3}$ & \cite{Bernt2020-ne}\\

$\rho_{M,an}$ & Catalyst density anode & \si[per-mode=fraction]{\gram\per\centi\meter\cubed} & 11.66 & \cite{Bernt2020-ne}\\

$\rho_{M,cat}$ & Catalyst density cathode & \si[per-mode=fraction]{\gram\per\centi\meter\cubed} & 21.45 & \cite{Bernt2020-ne}\\

$d_{M,an}$ & Catalyst crystal diameter anode & \si[per-mode=fraction]{\centi\meter} & $2.9\times10^{-7}$ & \cite{Bernt2020-ne}\\

$d_{M,cat}$ & Catalyst crystal diameter cathode & \si[per-mode=fraction]{\centi\meter} & $2.2\times10^{-7}$ & \cite{Bernt2020-ne}\\

$\lambda_{m}$ & Hydration factor of PEM & $\frac{\text{mol}_{\ch{H2O}}}{\text{mol}_{\ch{SO3}}}$  & 21 & \cite{Gorgun2006-ej}, Equation \ref{hyd_fact}\\

$\sigma_{mem}$ & Conductance of membrane & - & - & \cite{Gorgun2006-ej}, Equation \ref{conduct}\\

$p_{elec}$ & Price of electricity & - & - & \cite{noauthor_undated-so,Gagnon_undated-so}\\
$p_{water}$ & Price of water & $\frac{\$}{1000 \text{ gal}}$ & 2.78 & \cite{Water_Innovations_Inc_undated-io}\\

\bottomrule
\end{tabular}
}
\label{parameter nomenclature}
\end{table}

\begin{table}
\caption{Variable Nomenclature}
\resizebox{\columnwidth}{!}{%
\centering
\begin{tabular}{llcccc}
\toprule
Name & Description & Units & Value & Primary Decision Variable &Notes\\ \midrule

 $i(t)$ & Current density & \si{\ampere\per\square\centi\meter} & $0.1 - 4$& $\checkmark$ & - \\

 $\dot{n}_{\ch{H2O},1}(t)$ & Inlet water flowrate & \si{\mol\per\second} & $0 - 50,000$& $\checkmark$ & base case\\

  $\dot{n}_{3}(t)$ & Total flow out of anode & \si{\mol\per\second} & $0 - 10 N_c \frac{A_{cell}i}{4F}$& $\checkmark$ & base case\\

  $\dot{n}_{4}(t)$ & Total flow out of cathode & \si{\mol\per\second} & $0 - 10 N_c \frac{A_{cell}i}{2F}$& $\checkmark$ & base case\\

  $\dot{n}_{13}(t), \dot{n}_{14}$ & Split between satisfying demand and storage & \si{\mol\per\second} & $0 - 10^6$& $\checkmark$ & base case\\

  $\dot{n}_{16}(t)$ & Flowrate out of storage & \si{\mol\per\second} & $0 - 10^6$& $\checkmark$ & base case\\

  $\dot{n}_{11}(t)$ & Flowrate of \ch{N2} & \si{\mol\per\second} & $0 - 10^3$& $\checkmark$ & base case\\
  
 $N_{i}(t)$ & Moles of species $i$ & \si{\mol} & - & & -\\

$\dot{n}_{i,j}(t)$ & Molar flow rate of species $j$ in stream $i$ & \si[per-mode=fraction]{\mol\per\second} & - & & -\\

$y_{j,k}(t)$ & Mole fraction of species $j$ in stream at electrode $k$ & - & - & & -\\

$T(t)$ & Temperature of anode and cathode & \si{\degreeCelsius} & $60-80$ &  &base case\\

&  &  & $60-90$ & &sensitivity\\

$X_{\ch{H2}}$ & Conversion of hydrogen crossing from cathode to anode & - & 0.9 & &base case\\

$V^{total}(t)$ & Cell voltage & \si{\volt} & - &  &base case\\

$V^{undeg}_{d}(t)$ & Cell voltage not including degradation on day $d$ & \si{\volt} & - &  &base case\\

$\delta V^{deg}_{r}(t)$ & Intra-day change in degradation up to time $t$ & \si{\volt} & - &  &base case\\

$V^{cuml,deg}_{d}(t)$ & Cumulative degradation up to day $d$ and time $t$ & \si{\volt} & - &  &base case\\

$C^{elec}$ & Annual cost of electricity & \$ & - &  &base case\\

$C^{BoP,elec}$ & Annual cost of electricity used by balance of plant & \$ & - &  &base case\\

$C^{water}$ & Annual cost of de-ionized water & \$ & - &  &base case\\

$C^{\ch{N2}}$ & Annual cost of purge nitrogen & \$ & - &  &base case\\

$\delta_{r}$ & Change in storage over representative period $r$ & \si{\mole} & - &  &\\

$\Lambda_{\ch{H2}}$ & Hydrogen storage & \si{\mole} & - &  &\\
\bottomrule
\end{tabular}
}
\label{variablenomenclature}
\end{table}

\subsubsection{PEM Electrolyzer Potentials}

The current and voltage relationship used in this work borrows from the work done by \cite{Gorgun2006-ej,Liso2018-iy,Espinosa-Lopez2018-hb} and follows a standard procedure of estimating open circuit voltage and related overpotentials for the system as seen in Equation \ref{totvolt}. The total voltage is then the sum of these potentials.

\begin{align}
    V_{total} = V_{oc} + V_{act} + V_{ohm} + V_{deg} \label{totvolt}
\end{align}

Degradation of the cell appears as an increase in operating voltage of the cell. Discussion of the calculation of degradation for a cell as a function of usage is presented in the Optimization section.

\paragraph{Thermodynamic Potential}

Thermodynamic voltage is described by the free energy of reaction as shown in Equation \ref{eqvolt}. In general, this free energy of reaction as well as the entropy and enthalpy are functions of pressure and temperature. 

\begin{center}
    \begin{align}
            V_{rev} = \frac{\Delta G}{2F} \label{eqvolt}\\ 
            \Delta G = \Delta H - T\Delta S \label{gibbs}
    \end{align}
\end{center}

For this work, the enthalpy and entropy are assumed to be functions of temperature only. Correlations used to estimate enthalpies and entropies for \ch{H2O}, \ch{H2}, and \ch{O2} were taken from the NIST webbook \cite{BurgessUnknown-hi}. 

The Nernst Equation corrects this thermodynamic voltage for conditions other than the standard condition.

\begin{center}
    \begin{align}
        V_{oc} &= V_{rev}^{\circ} - \frac{RT}{2F}\ln\left( \frac{a_{H_{2}O}}{a_{H_{2}}a_{O_{2}}^{1/2}}\right) \label{Nernst Equation}\\
        V_{oc} &= V_{rev}^{\circ} + \frac{RT}{2F}\ln\left( \frac{P_{H_{2}}}{P^{\circ}}\sqrt{\frac{P_{O_{2}}}{P^{\circ}}}\right) \label{NernstSimple}
    \end{align}
\end{center}

As is consistent with other PEM electrolyzers models, the activity of water is assumed to be unity in order to reduce Equation \ref{Nernst Equation} to Equation \ref{NernstSimple} \cite{Gorgun2006-ej,Liso2018-iy,Espinosa-Lopez2018-hb}. The reference pressure is assumed to be 1 bar with $P_{\ch{H2}}$ and $P_{\ch{O2}}$ being 30 bar and 1 bar, respectively.

\paragraph{Activation Overpotential}
Activation overpotential is determined using the Butler-Volmer equation applied to the anode and cathode. The sum of the resulting overpotentials is the total activation overpotential.

\begin{center}
    \begin{align}
        V_{act} = \frac{RT}{\alpha_{an} F}\text{arcsinh}\left(\frac{i}{2i_{0,an}}\right) + \frac{RT}{\alpha_{cat} F}\text{arcsinh}\left(\frac{i}{2i_{0,cat}}\right)
    \end{align}
\end{center}

According to the kinetic theory used to derive the Butler-Volmer equation above, $\alpha_{an}$ and $\alpha_{cat}$ should sum to the number of electrons transferred in the reaction \cite{Thomas_Fuller_undated-ig}. As such, for a PEM, system they should sum to 2. However, in literature, $\alpha_{an}$, $\alpha_{cat}, i_{0,an},$ and $i_{0,cat}$ are used as fitting parameters for models when data for a polarization curve is available. This had led to a wide variety of charge transfer coefficients and exchange current densities. For this model charge transfer coefficients were left as fitting parameters for the system with the constraint that their sum could not exceed $n = 2$. Parameters were found as minimization of the sum of squared errors. Reference exchange current densities were chosen from typical values used in literature and corrected for temperature and roughness of the catalyst by Equations \ref{correctT} - \ref{correctrough} as was done by Görgün et al \cite{Falcao2020-vj, Gorgun2006-ej}.

\begin{center}
    \begin{align}
    i_{0,k}^{*} &= i_{0,ref,k}e^{\left[\frac{-E_{a}}{R}\left(\frac{1}{T}-\frac{1}{T_{ref}}\right)\right]} \label{correctT}\\
    \gamma_{M,k} &= \varphi_{l}m_{M,k}\frac{6}{\rho_{M,k}d_{M,k}}  \label{roughness}\\
    i_{0,k} &= \gamma_{M,k}i_{0,k}^{*} \label{correctrough}
    \end{align} 
\end{center}

\paragraph{Ohmic Overpotential} 
The ohmic resistance is a function of the thickness of the membrane, membrane conductivity, and operating current density. Membrane conductivity is a function of membrane hydration and operating temperature of the electrolyzer as seen in Equations \ref{ohm} - \ref{conduct}. Görgün et al used an empirical relation for the water hydration factor, $\lambda_{m}$, and $\sigma_{mem}$ where $\frac{1}{\sigma_{mem}} = R_{ohm}$ \cite{Gorgun2006-ej}. Under normal operating conditions the activity of water is such that $\lambda_m \approx 21$ , which was the value used for this study \cite{Maier_undated-zb}.

\begin{align}
    V_{ohm} &= \frac{\delta_{mem}}{\sigma_{mem}}i \label{ohm}
\end{align}

\begin{align}
    \lambda_{m} &= 0.43 + 17.81a_{H_{2}O} -39.85a_{H_{2}O}^{2} + 36a_{H_{2}O}^{3} \label{hyd_fact}\\
    \sigma_{mem} &= (0.00514\lambda_{m}-0.00326)e^{\left( 1268\left(\frac{1}{303}-\frac{1}{T}\right)\right)} \label{conduct}
\end{align} 

\subsubsection{PEM Electrolyzer Mass Balances}
\paragraph{Anode}\mbox{}\\
A transient water balance around the electrolyzer is shown in Equation \ref{wateranmol1}.

\begin{center}
    \begin{align}
        & \frac{dN_{H_{2}O,an}}{dt} = \dot{n}_{H_{2}O,2} - \dot{n}_{H_{2}O,3} - \dot{n}_{H_{2}O,3}^{vap} - \dot{n}_{H_{2}O,cross} - \dot{n}_{H_{2}O,consum} + \dot{n}_{H_{2}O,gen}\label{wateranmol1}\\
        & 0 = \dot{n}_{H_{2}O,2} - \dot{n}_{H_{2}O,3} - \dot{n}_{H_{2}O,3}^{vap} - \dot{n}_{H_{2}O,cross} - \dot{n}_{H_{2}O,consum} +\dot{n}_{H_{2},cross}X_{H_{2}}\label{wateranmol2}
    \end{align}
\end{center}

As is done on the industrial scale, the system is fed with excess water. As such, flows of water in and out of the anode are much greater than any liquid and vapor holdup in the anode. Therefore to simplify the system, the derivative term is assumed to be approximately zero in Equation \ref{wateranmol2}. It is also assumed there is a recombination catalyst embedded in the anode side of the membrane which has a 90\% conversion of \ch{H2} and \ch{O2} to \ch{H2O}. Conversion of hydrogen is represented by $X_{\ch{H2}}$. This recombination catalyst is a standard part of industrially operating PEM electrolyzers and are used to reduce the chance of a safety event occurring at the anode \cite{Christopher_Bryce_Capuano_Morgan_Elizabeth_Pertosos_Nemanja_Danilovic2018-lw,Nicolas_Guillet2014-tq}.

$\dot{n}_{H_{2}O,2}$ represents the liquid flowrate of water into the anode, and $\dot{n}_{H_{2}O,3}$ represents the liquid flowrate of water out of the anode. Other terms in the steady state water balance in Equation \ref{wateranmol2}, are defined below:

\begin{center}
    \begin{align}
        & \dot{n}_{H_{2}O,3}^{vap} = \frac{P_{H_{2}O}^{sat}}{P_{an}}\dot{n}_{gas,3}\label{watervap3}\\ 
        & \dot{n}_{H_{2}O,cross} = N_c\frac{n_g i A}{2F}\label{watercross}\\ 
        & n_g = 2.27 -  0.70i - 0.02P + 0.02Pi + 0.003T + 0.005iT - 0.0002PT\label{waterng} \\ 
        & \dot{n}_{H_{2}O,gen} = N_c \frac{iA}{2F} \label{waterconsum}
    \end{align}
\end{center}

In Equation \ref{watervap3}, water vapor is assumed to be in equilibrium with liquid water at the temperature of the cell in accordance with Raoult's Law. Equations \ref{watercross} and \ref{waterng} combined describe the electro-osmotic drift of water from anode to cathode at a given temperature, cathode pressure, and operating current density as described by Medina et al. \cite{Medina2010-lh}. Equation \ref{waterconsum} describes the consumption of water by Faraday's law assuming 100\% faradaic efficiency. 

Balances for oxygen and hydrogen at the anode are shown below in Equations \ref{oxymolbal1}-\ref{hydmolbal2}. Each species is based on a mole balance with Faraday's law of electrolysis modeling the electrochemical reaction term.

\begin{center}
    \begin{align}
        \frac{dN_{O_{2},an}}{dt} & = \dot{n}_{O_{2},gen} - \dot{n}_{O_{2},3} - \frac{1}{2}\dot{n}_{H_{2},cross}X_{H_{2}} \label{oxymolbal1}\\
        & =  N_{c}\frac{iA_{cell}}{4F} -y_{O_{2},an}\dot{n}_{gas,3} - \frac{1}{2}\dot{n}_{H_{2},cross}X_{H_{2}}  \label{oxymolbal2}\\
        \frac{dN_{H_{2},an}}{dt} &= \dot{n}_{H_{2},cross}(1-X_{H_{2}}) -  \dot{n}_{H_{2},out} \label{hydmolbal1}\\
        & = 0.31 \frac{i A_{cell}}{2F}(P_{\ch{H2},cat}-P_{\ch{H2},an})(1-X_{H_{2}}) - y_{H_{2},an}\dot{n}_{gas,3} \label{hydmolbal2}
    \end{align}
\end{center}

Equation \ref{oxymolbal1} represents the transient mole balance on oxygen. It is assumed the oxygen concentration of the water entering the anode is near zero, hence there is no oxygen flow into the anode. Equation \ref{oxymolbal2} is the same transient mole balance rearranged assuming oxygen is an ideal gas.

Equation \ref{hydmolbal1} represents the transient mole balance on hydrogen. The only way hydrogen can enter the anode is through crossover from the cathode, and the only way hydrogen can exit is through the anode exit gas stream. Equation \ref{hydmolbal2} is the same transient mole balance rearranged assuming hydrogen as an ideal gas and using the hydrogen crossover correlation investigated by Bernt et. al \cite{Bernt2020-af}.

\paragraph{Cathode}\mbox{}\\
A transient water balance on the cathode with the same holdup assumptions as the anode is shown in Equations \ref{watercatmol1} - \ref{watercatmol2}

\begin{center}
    \begin{align}
        & \frac{dN_{H_{2}O,cat}}{dt} = \dot{n}_{H_{2}O,cross} - \dot{n}_{H_{2}O,4} - \dot{n}_{H_{2}O,4}^{vap} \label{watercatmol1}\\
        & 0 = \dot{n}_{H_{2}O,cross} - \dot{n}_{H_{2}O,4} - \dot{n}_{H_{2}O,4}^{vap} \label{watercatmol2}
    \end{align}
\end{center}

The material balance for hydrogen and at the cathode are shown below in Equations \ref{hydcat1} - \ref{hydcat2}. Each species is based on a mole balance with Faraday's law of electrolysis modeling the electrochemical reaction term and gases are assumed ideal.

\begin{center}
    \begin{align}
            & \frac{dN_{H_{2}}}{dt} = \dot{n}_{H_{2},gen} - \dot{n}_{H_{2},4} - \dot{n}_{H_{2},cross} \label{hydcat1} \\
            & \frac{dN_{H_{2}}}{dt} = N_c\frac{iA}{2F} - y_{H_{2}}\dot{n}_{gas,4} - 0.31 \frac{i A_{cell}}{2F}(P_{\ch{H2},cat}-P_{\ch{H2},an})(1-X_{H_{2}}) \label{hydcat2}
    \end{align}
\end{center}

Oxygen crossover is assumed negligible as the oxygen is much less permeable to the Nafion membrane, and the cathode is pressurized meaning oxygen transport will be effectively zero \cite{Bernt2020-af}.

\subsubsection{PEM Electrolyzer Energy Balance}
The energy balance is based on a lumped capacitance model described in Equations \ref{energybalance}-\ref{energyloss}. Water enters the anode at ambient conditions so that $T_{in} = T_{ref}$. Espinosa-Lopez et. al presented a thermal lumped capacitance model for a 60 cell PEM electrolyzer stack in which they reported a lumped thermal capacitance $C_{p,th}$ and thermal resistance $R_{th}$ normalized by the area of the cell \cite{Espinosa-Lopez2018-hb}. These values for the lumped thermal capacitance and heat loss resistance were used in the model as a reference and scaled up or down based on the area of the modeled system. $V_{th}$ is the thermoneutral voltage of the cell, assumed to be 1.48 \si{\volt}.

\begin{align}
    \frac{dT}{dt} &= \frac{1}{C_{p,th}}\left(\dot{H}_{in}-\dot{H}_{out} +\dot{Q}_{gen} - \dot{Q}_{loss}\right) \label{energybalance}\\
    \dot{W}_{in} &= \sum_{i} n_{i,in}H_i(T_{in}) \label{energyin}\\
    \dot{W}_{out} &= \sum_{i} n_{i,out}H_i(T_{cell}) \label{energyout}\\
    \dot{Q}_{gen} &= N_{c}\left(V_{total} - V_{tn}\right)iA \label{energygenerated}\\
    \dot{Q}_{loss} &= \frac{1}{R_{th}}\left(T-T_{ambient}\right) \label{energyloss}\\
\end{align}

The standard enthalpies of formation are included in the thermoneutral voltage, and $T_{ref} = T_{in} = T_{ambient}$ for the thermodynamic calculations. This is consistent with the energy balance presented in Espinosa et al and allows the  heat generation term to be written as the potential applied over the thermoneutral voltage \cite{Espinosa-Lopez2018-hb}.

Equation \ref{energybalance} represents the overall energy balance for the system including all possible sources and sinks for heat. Equations \ref{energyin} - \ref{energyloss} define those sinks and sources. For the energy in, species include water and nitrogen (if used), and for the energy out species include,  water, water vapor, nitrogen (if used), hydrogen, and oxygen.

\subsection{K-Means Clustering for Representative Days} \label{clustering description}

K-means clustering was performed on the input price data series of hourly electricity prices. The k-means clustering algorithm works to find $k$ centroids to which each of the 365 days in the year most closely correspond. 

Once each day is assigned a cluster, the day closest to the cluster centroid is chosen as the representative day for that cluster. Each of these representative days then have an associated weight which signifies the number of real days that belong to the cluster. This way, the optimization can be performed in a significantly reduced time domain while still capturing much of the variability in electricity price seen throughout the year.

\FloatBarrier
\subsection{Storage Formulation} \label{storageformulation}

\FloatBarrier
\subsection{Convergence Time}

\begin{table}[!h]
\caption{Convergence time for scenarios presented in the paper. Each case took 15 iterations of the GSS in order to converge within 0.1\% tolerance.}
\centering
\begin{tabular}{lc}
\toprule
Case                       & Time (hours) \\ \midrule
Degradation 2022 Base Case & 44.8         \\
No Degradation             & 25.2         \\
Degradation, West Texas    & 74.3         \\
Degradation, Fixed CAPEX   & --           \\
High Temperature           & 57.6         \\
No Safety Constraint, Fixed CAPEX       & --         \\
Degradation 2030 Base Case & 52.1         \\
2030 High CAPEX            & 48.4         \\
2030 Low CAPEX             & 43.7         \\
2030 High Decarbonization  & 45.7         \\
2030 Low Decarbonization   & 49.4           \\ \bottomrule
\end{tabular}
\label{convergence_time}
\end{table}

\FloatBarrier
\subsection{Algorithm Scaling}

Convergence time increases with an increasing number of representative days. The vast majority of constraints for each problem type are equality constraints. The LCOH seems to be converging to a singular value, albeit non monotonically. Theoretically, as the number of representative days approaches the number of real days, the LCOH would approach the real LCOH which is the LCOH of using the full price series data. 7 representative days was chosen for this study for the base case's ability to converge $< 48$ hours. For this particular price series it happens to be the number of representative days that has the lowest LCOH, though it was not chosen for that reason.

\begin{table}[!h]
\caption{Convergence time for different numbers of representative days. Each case was run with the base case South LZ price series with degradation.}
\centering
\begin{tabular}{lccccccc}
\toprule
\thead[l]{Representative \\ Days} & \thead[c]{LCOH \\ (\$/kg)}   & \thead[c]{Convergence \\Time \\ (hours)}  &\thead[c]{Number of \\ Variables} & \thead[c]{Number of \\Equality \\Constraints} & \thead[c]{Number of \\ Inequality\\ Constraints} &\thead[c]{Number of GSS \\ Iterations} &\thead[c]{Avg. GSS \\ Iteration \\ Time \\(min)}\\
\midrule
2                   & \$                       6.96 & 23.3                     & 188,469             & 186,728                        & 776    & 15 & 21.8                          \\
4                   & \$                       6.92 & 31.5                     & 199,913             & 196,796                        & 1,552  & 15 & 29.5                          \\
7                   & \$                       6.60 & 44.8                     & 217,079             & 211,898                        & 2,716    & 15 & 39.6                        \\
9                   &\$ 6.83                               &   49.4               &    228,523    &        221,966             &               3,492  & 15 & 46.3                                                 \\
10                  & \$                 7.06             &    57.5                      &      234,245               &      227,000                          &   3,880   & 15 & 53.9                            \\
14                  &      \$ 7.19                         &   82.3                       &     257,133                &       247,136                         &  5,432 & 15 & 77.2\\ \bottomrule                              
\end{tabular}
\label{algo_scaling}
\end{table}

\FloatBarrier
\subsection{Results from Alternate Scenarios}

\subsubsection{Degradation Sensitivities}

\FloatBarrier
\subsubsection{Degradation with CAPEX Fixed at No Degradation Result}

\FloatBarrier
\subsubsection{Impact of Higher Temperature}

\FloatBarrier
\subsubsection{Impact of Recombination Catalyst and Safety Constraint}

\FloatBarrier
\subsubsection{Future Scenarios}

\begin{figure}
     \centering
     \includegraphics[width=\textwidth]{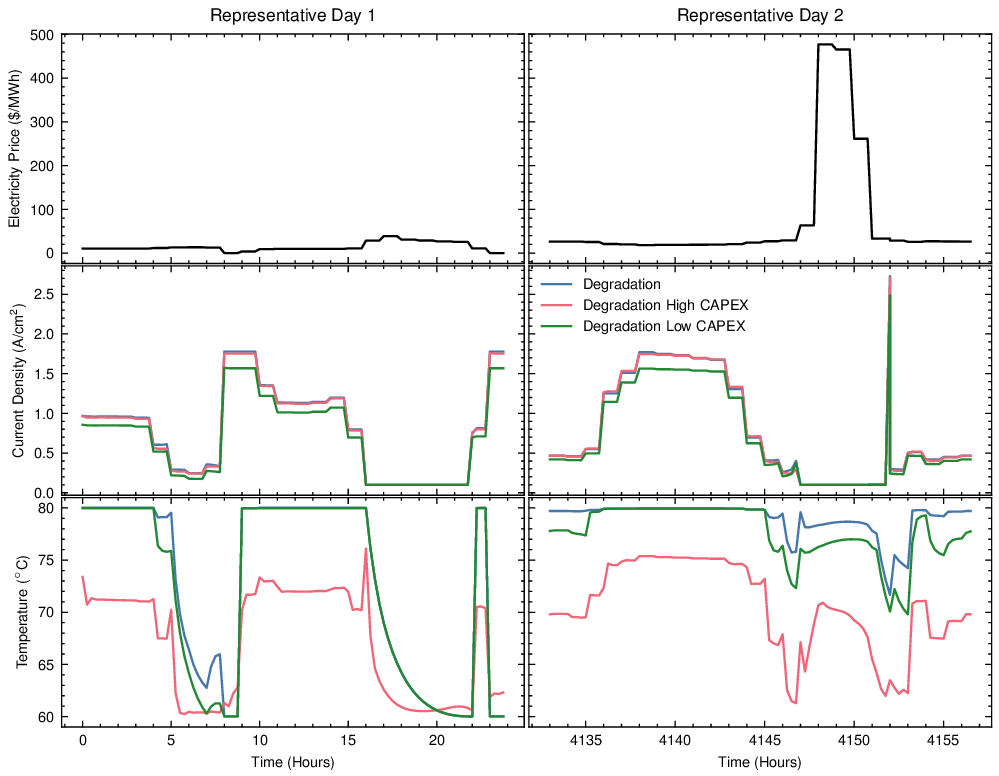} 
     \caption{Temperature profile of future CAPEX cases. Generally, the high CAPEX case with more cells operates at lower temperatures due to lower current density and therefore lower overpotential.}
     \label{Temp_Only}
\end{figure}

\begin{figure}[!ht]
     \centering
     \begin{minipage}[c]{\textwidth}
     \includegraphics[width=\textwidth]{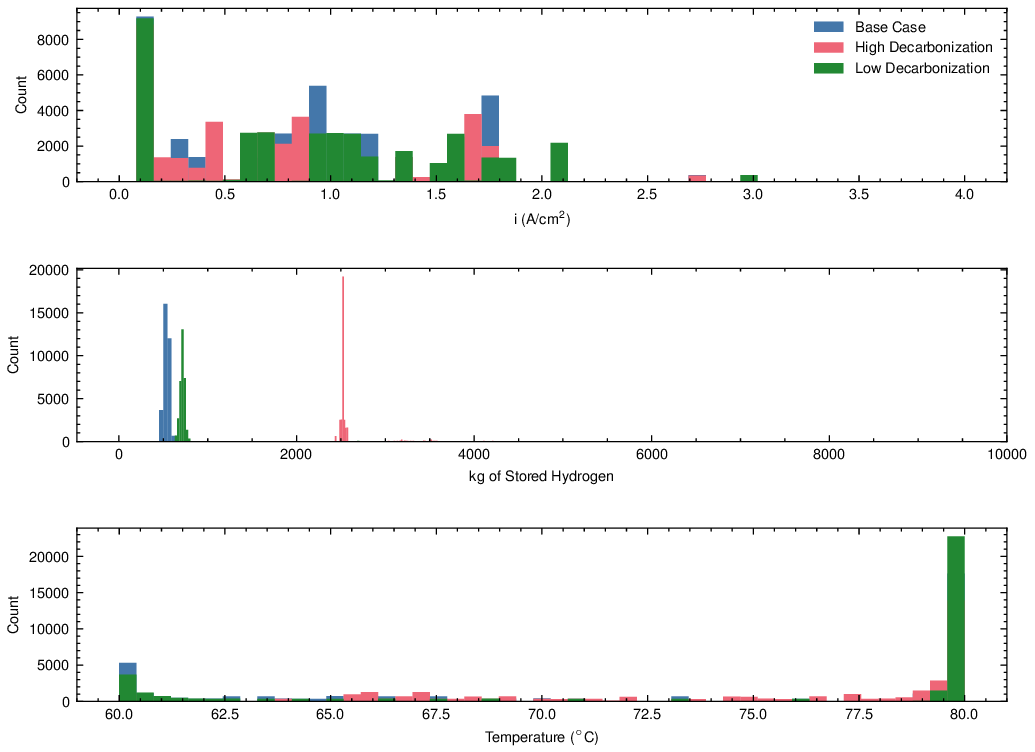} 
     \caption{Distribution of current density, hydrogen storage, and temperature for the future cases run with low, mid, and high decarbonized grids. Price scenario corresponding to high grid decarbonization leads to sizing greater storage and operating at lower current densities. Lower grid decarbonization based price scenarios allows for operation at higher current densities because of the lower price of electricity. However, these effects largely do not affect the LCOH (see Table \ref{results}).}
     \label{decarb}
     \end{minipage}
\end{figure}

\clearpage

\printbibliography
\end{refsection}

\end{document}